\documentclass[aps, prd, 10pt, twocolumn, superscriptaddress,noshowpacs, preprintnumbers, longbibliography,nofootinbib,bibnotes,hyperref,floatfix]{revtex4-2}
\usepackage{graphicx}
\usepackage[dvipsnames]{xcolor}
\usepackage{amsmath,amssymb,bigints}

\usepackage{tikz,xcolor,hyperref}

\definecolor{lime}{HTML}{A6CE39}
\DeclareRobustCommand{\orcidicon}{\hspace{-1mm}
	\begin{tikzpicture}
	\draw[lime, fill=lime] (0,0) 
	circle [radius=0.16] 
	node[white] {{\fontfamily{qag}\selectfont \tiny \,ID}};
	\draw[white, fill=white] (-0.0525,0.095) 
	circle [radius=0.007];
	\end{tikzpicture}
	\hspace{-3mm}
}

\foreach \x in {A, ..., Z}{\expandafter\xdef\csname orcid\x\endcsname{\noexpand\href{https://orcid.org/\csname orcidauthor\x\endcsname}
			{\noexpand\orcidicon}}
}

\begin{document}

\title{Diffuse Neutrino Background from Magnetorotational Stellar Core Collapses}

\author{Pablo Mart\'{i}nez-Mirav\'e\orcidA{}}
\email{pablo.mirave@nbi.ku.dk}
\affiliation{Niels Bohr International Academy and DARK, Niels Bohr Institute, University of Copenhagen, \\
Blegdamsvej 17, 2100, Copenhagen, Denmark} 

\author{Irene Tamborra\orcidB{}}
\email{tamborra@nbi.ku.dk}
\affiliation{Niels Bohr International Academy and DARK, Niels Bohr Institute, University of Copenhagen, \\
Blegdamsvej 17, 2100, Copenhagen, Denmark}

\author{Miguel \'{Angel} Aloy\orcidD{}}
\email{miguel.a.aloy@uv.es}
\affiliation{Departament d’Astronomia i Astrof\'{i}sica, Universitat de Val{\`e}ncia, Dr.~Moliner, 50, 46100, Burjassot, Spain}
\affiliation{Observatori Astron{\`o}mic, Universitat de Val{\`e}ncia, 46980, Paterna, Spain}

\author{Martin Obergaulinger\orcidC{}}
\email{martin.obergaulinger@uv.es}
\affiliation{Departament d’Astronomia i Astrof\'{i}sica, Universitat de Val{\`e}ncia, Dr.~Moliner, 50, 46100, Burjassot, Spain}

\begin{abstract}
A statistically significant detection of the diffuse supernova neutrino background (DSNB) is around the corner.
To this purpose, we assess the contribution to the DSNB of  magnetorotational collapses of massive stars, relying on a suite of state-of-the-art three-dimensional neutrino-magnetohydrodynamic simulations. We  find that neutrinos from magnetorotational core collapses boost the high-energy tail of the DSNB spectrum, similar to what is expected from neutrino-driven black hole-forming collapses.
The latest data from the  Super-Kamiokande Collaboration can already exclude that more than $9\%$ of all collapsing massive stars undergo magnetorotational collapses under optimistic assumptions.  A DSNB detection at $3 \sigma$ could take place up to $4$~yr earlier at Super-Kamiokande-Gadolinium or JUNO if the fraction of magnetorotational collapses  should be larger than $10\%$.
Fascinatingly, if the fraction of magnetorotational stellar collapses should be larger than $16\%$ ($11\%$), Hyper-Kamiokande could measure such a fraction at $3\sigma$ after ($10$~yr) $20$~yr of DSNB data taking.
The combination of DSNB and electromagnetic data has the potential to resolve  the degenerate contributions  from magnetorotational and  neutrino-driven black hole-forming collapses, providing crucial insight on the properties of the population of collapsing massive stars.

\end{abstract}

\maketitle

\section{Introduction}

On average, every second, a core-collapse supernova (CCSN) explodes somewhere in the Universe releasing \mbox{$\sim 10^{53}$~erg} in the form of neutrinos and antineutrinos of all flavors. The cumulative flux of  neutrinos emitted from all CCSNe in the Universe over cosmological distances is the Diffuse Supernova Neutrino Background (DSNB)~\cite{Bisnovatyi-Kogan:1982oyy,Krauss:1983zn,1984SvA....28...30D,Beacom:2010kk,Lunardini:2010ab,Mirizzi:2015eza,Ando:2023fcc}. The DSNB is expected to be the dominant contribution to the overall diffuse emission of neutrinos from known sources between $10$ and $25$~MeV~\cite{Vitagliano:2019yzm}.  

To date, the theoretical modeling of the DSNB is hampered by our incomplete understanding of the CCSN  rate~\cite{Ekanger:2023qzw} and the initial mass function~\cite{Ziegler:2022ivq}, the rate of collapses forming a black hole (BH) ~\cite{Lunardini:2009ya,Nakazato:2024gem}, flavor conversion in the supernova (SN) core~\cite{Lunardini:2012ne},  the metallicity evolution of  galaxies~\cite{Nakazato:2015rya,Ashida:2023heb}, binary stellar interactions~\cite{Horiuchi:2020jnc}, as well as the neutron star properties~\cite{Kresse:2020nto}; we refer the interested reader to Ref.~\cite{Kresse:2020nto} for an overview of the relevant uncertainties. 
Moreover, Type Ia SNe~\cite{Anandagoda:2023sbg}, population III supermassive stars~\cite{Nakazato:2005ek,Suwa:2008sf,Nagele:2021kvw}, and accretion flows~\cite{Schilbach:2018bsg,Wei:2024qgy} can also contribute to the diffuse neutrino emission in the same energy region.  

The DSNB is the key observational target of the  Super-Kamiokande-Gadolinium project~\cite{Super-Kamiokande:2021the,Super-Kamiokande:2023xup}. While the DSNB signal has not been detected with high confidence yet, the first results from the Super-Kamiokande-Gadolinium project are extremely encouraging~\cite{Super-Kamiokande:2021the,Super-Kamiokande:2023xup,Super-Kamiokande:2024kcb,SK-Gd:2024}. 

So far, most  DSNB searches have focused on the detection of electron antineutrinos via inverse beta decay (i.e., $\bar{\nu}_e + p \to e^+ + n$). Whereas this approach remains the most promising one~\cite{Beacom:2003nk,Super-Kamiokande:2021the,Super-Kamiokande:2024kcb,SK-Gd:2024}, a measurement of the DSNB in all flavors, and for neutrino and antineutrinos, is desirable. In this regard, the liquid Argon neutrino telescope DUNE should improve existing limits on $\nu_e$~\cite{Moller:2018kpn,DUNE:2020lwj}; detectors based on the neutrino-nucleus coherent elastic scattering, such as DARWIN and RES-NOVA, could probe the non-electron flavors~\cite{Suliga:2021hek,Lang:2016zhv,Pattavina:2020cqc}.

In this paper, we investigate the contribution to the diffuse neutrino emission from magnetorotational core collapses of massive stars, i.e. from the collapse of rapidly rotating stars for which the rotation and magnetic fields are relevant. We differentiate between two magnetorotational collapse scenarios according to the nature of their remnant: a protomagnetar or a spinar. We refer to rapidly rotating protoneutron stars with strong magnetic fields as protomagnetars. We adopt the term spinar for a fast rotating magnetized relativistic object, in quasi-equilibrium due to the balance between the centrifugal and gravitational forces, which undergoes a two-stage collapse, first to a protoneutron star and, subsequently, to form a BH~\cite{Lipunova:2009qf,1983A&A...127L...1L,1987ans..book.....L,1998A&A...329L..29L}~\footnote{The term ``spinar'' was first introduced in Ref.~\cite{1971swng.conf..485M} in the context of active galactic nuclei.}

Magnetorotational core collapses are considered to be the progenitors of superluminous SNe and certain gamma-ray bursts~\cite{Aguilera-Dena:2018ork,Obergaulinger:2021omt}. To this purpose, we model the neutrino emission from the (transiently stable) protomagnetars  relying on three-dimensional neutrino-magnetohydrodynamic simulations of the same models studied in 2D in Ref.~\cite{Obergaulinger:2021omt}. Since the rate of protomagnetars and spinars is unknown, we keep it as a free parameter and  investigate whether the neutrino emission from magnetorotational core collapses could lead to a DSNB  contribution which is  degenerate with respect to the uncertainties in the CCSN rate and fraction of neutrino-driven BH-forming collapses.

Our work is organized as follows. In Sec.~\ref{sec:models}, we introduce the suite of models adopted to compute the DSNB. Section~\ref{sec:dsnb} presents our fiducial DSNB model from  CCSNe and neutrino-driven BH-forming collapses as well as explore the contribution from protomagnetars and spinars.  Section~\ref{sec:current} presents a comparison between our theoretical prediction and the model-independent limits on the $\bar\nu_e$ flux from  Super-Kamiokande~\cite{SK-Gd:2024}. A large fraction of magnetorotational core collapses would enhance the near-future prospects for detecting the DSNB at Super-Kamiokande-Gadolinium and JUNO, as shown in Sec.~\ref{sec:future}. There, we also explore the sensitivity of Hyper-Kamiokande loaded with Gadolinium to measure the fraction of protomagnetars and spinars contributing to the DSNB. Section~\ref{sec:em}  highlights the relevance of combining the DSNB data with  electromagnetic observations to overcome the astrophysical uncertainties currently plaguing the DSNB. Finally, we conclude in Sec.~\ref{sec:conclusions}. 
\section{Neutrino driven and magnetorotational  core collapses: Neutrino emission properties \label{sec:models} }
In this section, we introduce the neutrino emission properties for successful neutrino-driven CCSNe  
and neutrino-driven BH-forming models, as well as the protomagnetar and spinar models adopted to reproduce the  population of collapsing massive stars. 

\subsection{Core-collapse supernova  and black hole forming collapse models}

In order to model the DSNB contribution from neutrino-driven core collapses, we follow Ref.~\cite{Moller:2018kpn} and
rely on  spherically symmetric (1D) hydrodynamical simulations without muons~\cite{Mirizzi:2015eza,Garching}. We consider a  model with a mass of  $11.2 M_\odot$ as representative of the neutrino-driven CCSN population with mass below $15 M_\odot$. For the remaining  neutrino-driven CCSN population, we use a  model with mass of $27 M_\odot$ and a neutrino-driven BH-forming  model with a mass of 40$M_\odot$ (model s4027b2, characterized by a short accretion phase, with the neutrino emission lasting for $0.57$~s after bounce before prompt BH formation). All models employ the Lattimer and Swesty equation of state, with a nuclear incompressibility modulus of $K = 220$~MeV~\cite{Lattimer:1991nc}.

\begin{figure*}
    \centering
    \includegraphics[width=\linewidth]{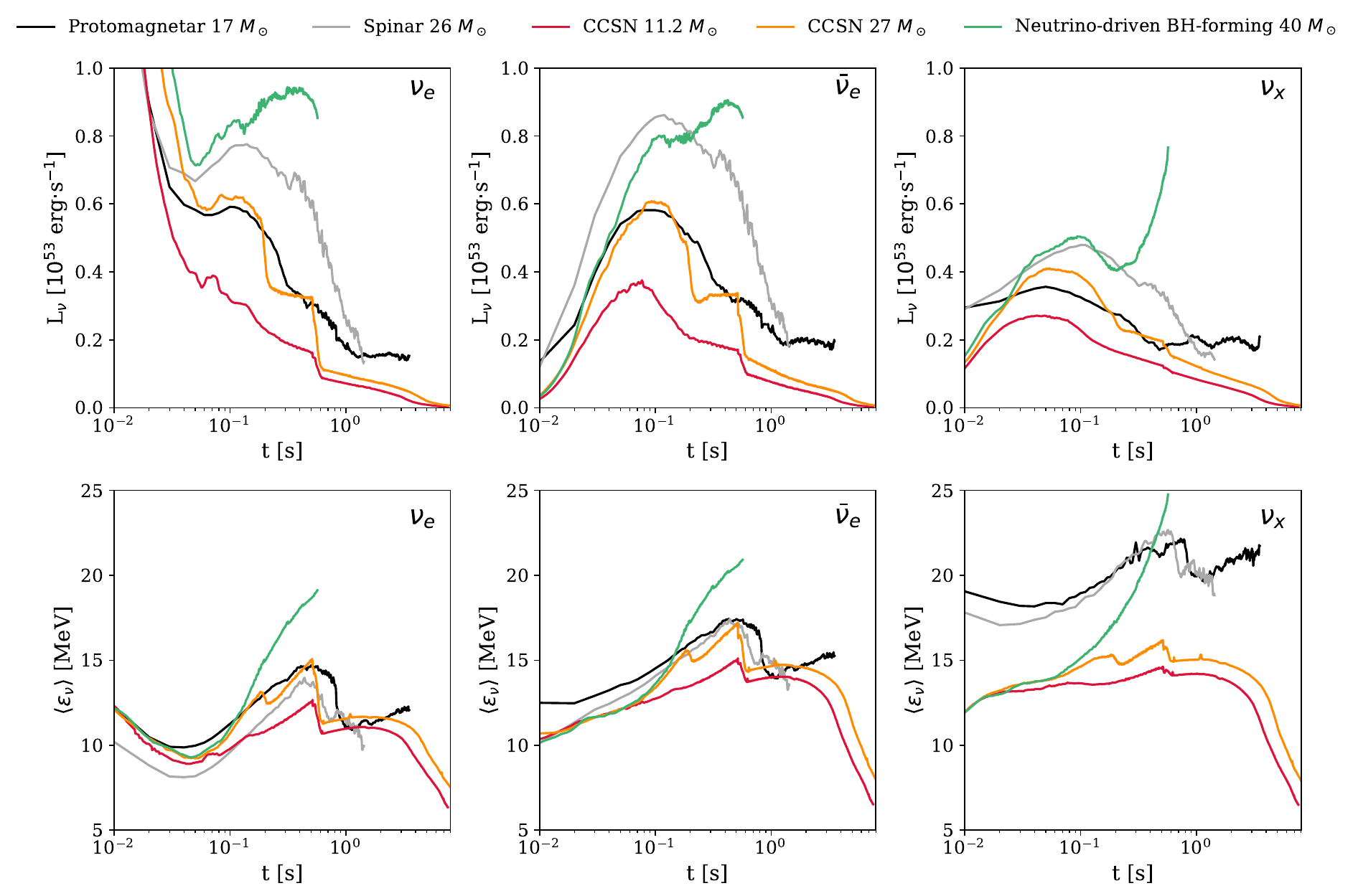}
    \caption{Temporal evolution of the neutrino emission properties for one representative protomagnetar model with mass of $17~M_\odot$ (1D equivalent neutrino emission properties of model A17, black lines), the  spinar model with mass 26 $M_\odot$ (1D equivalent neutrino emission properties of model  of model A26, gray lines), two 1D neutrino-driven CCSNe models with mass of $11.2~ M_\odot$ and $27~ M_\odot$ (red and orange lines, respectively), and a 1D neutrino-driven BH-forming collapse with $40~M_\odot$ (model  s4027b2 characterized by a short accretion phase before prompt BH formation, green lines). 
    The upper and lower panels display the luminosity and mean energy,  for $\nu_e$, $\bar\nu_e$, and $\nu_x$,
    respectively. The protomagnetar and spinar models exhibit larger average energies for the non-electron flavors due to neutrino decoupling occurring at higher densities than for the neutrino-driven CCSN models.
    }
    \label{fig:A17}
\end{figure*}

Figure~\ref{fig:A17} displays the temporal evolution of the neutrino emission properties (for $\nu_e$, $\bar\nu_e$, and $\nu_x$, where the latter  denotes  $\nu_\mu$, $\bar{\nu}_\mu$, $\nu_\tau$, and
$\bar{\nu}_\tau$) for the neutrino-driven collapse models introduced above. One can see that the duration of the neutrino signal is shorter for the neutrino-driven BH-forming model, with neutrino luminosity slightly larger than the one of the neutrino-driven CCSN models. 

\subsection{Protomagnetar and spinar models}
 In order to model the neutrino emission from protomagnetars we rely on 3D simulations combining  special relativistic magnetohydrodynamics and a spectral two-moment (M1) neutrino transport~\cite{Obergaulinger:2020cqq, Obergaulinger:2021omt}. The simulations use the SFHo hadronic equation of state~\cite{Steiner:2012rk} and include the most relevant neutrino-matter reactions (absorption, emission, and scattering by nucleons and nuclei, scattering off electrons and positrons, electron-positron pair annihilation and nucleonic bremsstrahlung).
The models  are the 3D equivalent of the  2D ones presented in Ref.~\cite{Obergaulinger:2021omt}.
We call protomagnetar a post-collapsed core that develops magnetar magnetic field strengths $\gtrsim \mathcal{O}(10^{15})$~G and has an associated magnetar phenomenology as described in, e.g., Refs.~\cite{Metzger_2011MNRAS.413.2031,Metzger_2018ApJ...857...95}; if the collapse into a BH occurs a few seconds after core bounce, we refer to our models as spinars.
Note that we do not consider the contribution to the neutrino background from the resulting  BH accretion disk, which was investigated in  Ref.~\cite{Schilbach:2018bsg}.

We rely on seven  models with mass at the zero age main sequence time: $M_{\mathrm{ZAMS}} = 5~M_\odot$, $8~M_\odot$, $13~M_\odot$, $17~M_\odot$, $20~M_\odot$, $26~M_\odot$, and $30~ M_{\odot}$, obtained from the progenitors presented in Ref.~\cite{Aguilera-Dena:2018ork}. 
These models  consider enhanced rotational mixing, leading to chemically homogeneous evolution. 
While a detailed description of the dynamics and explosion properties of these 3D models will be presented elsewhere, here we provide a brief summary. 

All cores produce delayed explosions at times between $\sim 0.3$ and $\sim 1.3$~s after bounce.  Shock revival is powered by a combination of neutrino heating and magnetorotational stresses, with a higher contribution of the latter leading to explosions with energies above $E_{\mathrm{exp}} \gtrsim 10^{51}$~erg. Unlike the 2D models presented in Ref.~\cite{Obergaulinger:2021omt}, only the 3D models with the larger masses ($M_\mathrm{ZAMS} \gtrsim 
26 M_\odot$; models A26 and A30 in the nomenclature of that reference) have clear signs of BH formation, leading to spinars. The formation of the BH does not happen promptly after core bounce, but, at least, after $1.5$~s. This is the case of model A26, for example. For more massive models, more than $2.5$~s are needed for the protomagnetars to collapse. A common feature of these high-mass models is that the protoneutron star mass grows (due to non-spherical accretion) above the limit set by the equation of state relatively soon after bounce. High-mass progenitors thus form protomagnetars for a relatively short period of time, and then they may turn into BH-forming events or even collapsars at later times, as anticipated in Refs.~\cite{Obergaulinger:2017qno,Aloy:2020svw}. Lower mass progenitors (including models A05, A08, A13, A17, and A20) do not display signatures of BH formation;  the protoneutron star mass remains below the maximum mass set by the equation of state throughout the  computed evolution (for some models, such as A17, the evolution lasts for more than $3.5$~s after bounce). These lower mass progenitor models are more likely to form standard protomagnetars that survive the SN explosion. It is noteworthy that the progenitor stars modeled in Ref.~\cite{Aguilera-Dena:2018ork}  were predicted to produce either superluminous SNe (if $M_\textrm{ZAMS}\lesssim 13 M_\odot$), protomagnetar-driven SNe ($M_\textrm{ZAMS}=17  M_\odot$, $20 M_\odot$), or collapsars ($M_\textrm{ZAMS}\gtrsim 26 M_\odot$). However, the limited simulated time for our 3D models does not allow to distinguish whether the protomagnetar models  lead to superluminous SNe or protomagnetar-driven explosions, so far. Nevertheless, the distribution of specific angular momentum in spinars (see, e.g., Fig.~2 of Ref.~\cite{Obergaulinger:2021omt}) suggests promising prospects for the formation of an accretion disc, and hence, the potential for a collapsar scenario.

The M1 neutrino transport evolves the spectral energy and momentum densities, $J (\vec{r}, E)$ and $\vec{H} (\vec{r}, E)$, of $\nu_e$, $\bar{\nu}_e$, and $\nu_x$ as functions of position ($\vec{r}\,$) and particle energy ($E$).  As the momentum density is equal to the energy flux of the neutrinos, we can extract the luminosity directly from its radial component by integrating it at an extraction radius, fixed at $r_{\mathrm{ext}} = 500$~km, over the solid angle: $L_{\nu} (E) = \int_{\Omega} \mathrm{d} \Omega \, r_{\mathrm{ext}}^2 J^r (r_{\mathrm{ext}},E)$.  The number density of the neutrinos and its flux can be obtained by dividing the spectral energy and momentum density by the particle energy, $\{N(\vec{x},E), \vec F_N(\vec{x},E)\} =  {1}/{E} \{J(\vec{x},E), \vec H(\vec{x},E)\}$.  We then compute the neutrino mean energies via an integration over the energy spectrum: $\langle \varepsilon_{\nu} \rangle  = \langle \int \mathrm{d}E \, J(\vec{x},E) \rangle_{\Omega,r_{\mathrm{ext}}} / \langle \int \mathrm{d}E \, N(\vec{x},E) \rangle_{\Omega,r_{\mathrm{ext}}} $, where $\langle ... \rangle_{\Omega,r_{\mathrm{ext}}}$ indicates an average over the solid angle at the extraction radius.

Figure~\ref{fig:A17} shows the angle-averaged (1D equivalent) neutrino emission properties for the representative protomagnetar  model with mass $M = 17 M_\odot$ (model A17; note that the sudden termination of the neutrino signal at about $4$~s corresponds to the end of the simulation time  for this model and not to BH formation) and the spinar model with mass 26~$M_\odot$ (model A26; the signal terminates about $1.5$~s after core bounce due to BH formation). 
These models have neutrino luminosity $\text{L}_\nu \gtrsim 10^{52}$~erg$\cdot$s$^{-1}$ for all flavors, with the typical neutrino emission lasting for several seconds. 
The neutrino mean energies for the different flavors exhibit a characteristic hierarchy: $\langle \varepsilon_{\nu_e} \rangle < \langle \varepsilon_{\bar{\nu}_e} \rangle < \langle \varepsilon_{\nu_x} \rangle$. The large mean energy of the non-electron flavor neutrinos (which exceeds $20$~MeV for a significant part of the emission) is due to the  compact protoneutron star, that facilitates the neutrino decoupling at  densities higher than in the neutrino-driven CCSN models. 
The small offset (of a few MeV) in the heavy-neutrino mean energies is produced by the development of steep density and temperature gradients at the typical densities where the transition region from diffusion to free streaming is located. As a result, the flux factors (i.e., the ratio of flux-to-neutrino energy) grow over a narrow radial range. The region where $\nu_x$'s decouple from matter and  the other neutrino flavors coincides with a sharp drop of the gas temperature, such that even a small radial difference has an impact on the mean  energies of the non-electron neutrino flavors. The detailed treatment of the pair processes may have also some (small) impact on the computed heavy-neutrino mean energy. 

\section{Diffuse neutrino background~\label{sec:dsnb}}
In this section, after outlining the approach employed to model the DSNB, we compute  the contribution of neutrino-driven CCSN explosions, neutrino-driven BH-forming collapses, protomagnetars, and spinars to the diffuse neutrino emission. 

\subsection{Contribution to the diffuse neutrino background from neutrino-driven core-collapse supernovae and black hole-forming collapses\label{subsec:dsnb-ccsn}}

The diffuse neutrino background from neutrino-driven CCSNe and neutrino-driven BH forming collapses is given by:
\begin{align}
    \Phi_\text{CCSN}(E) =  &c\int_{8 M_\odot}^{125 M_\odot} \textrm{d}M \nonumber \\
    &\,\times\int_{z = 0}^{z_\text{max}} \text{d}z\frac{\mathcal{R}(z, M)}{H(z)}F_\nu(M, E(1+z))\, ,
    \label{eq:ccsn-dsnb}
\end{align}
where $c$ denotes the speed of light, $z$ is the redshift with $z_\text{max}=5$ (see below), $\mathcal{R}(z, M)$ is the CCSN rate, $F_\nu(M, E(1+z))$ is the time-integrated neutrino energy  spectrum of a CCSN of mass $M$, and the factor $1/H(z)$ accounts for the expansion history of the Universe. 

\paragraph*{Neutrino energetics.}~The time-integrated neutrino energy spectrum, $F_\nu(M, E)$, is computed as
\begin{align}
    F_\nu(M,E) = \bigintssss \text{d}t\,\mathrm{L}_\nu(E, t)\frac{\phi^0_\nu(E,t)}{\langle \varepsilon_\nu(t) \rangle}\, ,
    \label{eq:emission-spec}
\end{align}
where we integrate over the duration of neutrino emission the product of the luminosity ($\mathrm{L}_\nu(E,t)$), the spectral energy distribution ($\phi^0_\nu(E,t)$  parametrized following Refs.~\cite{Keil:2002in,Tamborra:2012ac}), and the neutrino mean energy  ($\langle \varepsilon_\nu(t) \rangle$).

\begin{figure}
    \centering
    \includegraphics[width=\linewidth]{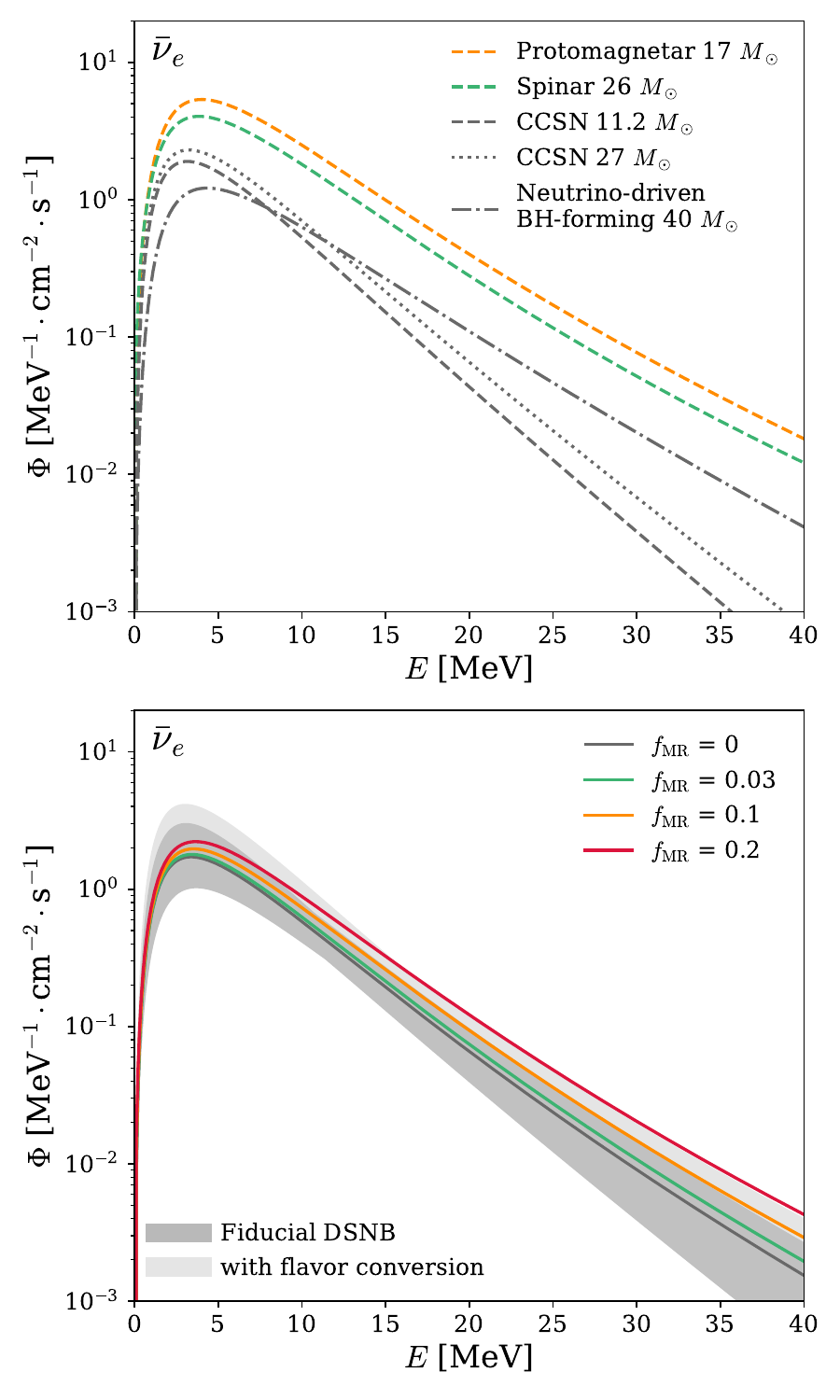}
    \caption{{\it Top panel}: Electron-antineutrino DSNB flux (Eq.~\ref{eq:ccsn-dsnb}) as a function of the observed neutrino energy, assuming that the $11.2 ~M_\odot$ or the $27~M_\odot$,  the $40~M_\odot$  model  (dashed, dotted, and dashed-dotted gray lines), the  protomagnetar model with  mass  of $17~M_\odot$ (dashed orange line), or the spinar model with mass of $26~M_\odot$ (dashed green line) are representative of the  population of collapsing massive stars, respectively. Flavor conversion at the source is neglected. {\it Bottom  panel}: Same as top panel, but now taking into account different CCSN and magnetorotational models according to the initial mass function (cf.~Fig.~\ref{fig:imf_che}) and  for different values of the fraction of magnetorotational collapses ($f_\text{MR}$, colored lines). The dark gray band shows the fiducial DSNB model and the light gray one takes into account  complete flavor conversion at the source. }
    \label{fig:model-prediction}
\end{figure}

The top panel of Fig.~\ref{fig:model-prediction} shows the $\bar\nu_e$ DSNB spectrum computed following Eq.~\ref{eq:ccsn-dsnb} and assuming that all collapsing massive stars behave as  the $11.2~M_\odot$, or $27~M_\odot$ CCSN models, or the $40~M_\odot$ neutrino-driven BH-forming model (i.e., we compute the DSNB  assuming that each of these models would be representative of the whole core collapse population in order to highlight the differences in the neutrino emission). One can see that, because of the  energetic and long-lasting emission from protomagnetars (cf.~$17~M_\odot$ model), their  flux is the largest with the  high-energy tail as soft as that of BH-forming CCSNe. Similarly, the luminous neutrino emission from spinars results in the second largest flux (after protomagnetars, cf.~$26 M_\odot$ model) with a  soft high-energy tail comparable to the one of protomagnetars.

\paragraph*{Core-collapse supernova rate.}~The  bulk of the DSNB mainly comes from neutrinos emitted from core collapses at $z \lesssim z_\text{max}$~\cite{Ando:2004hc}. For these redshifts, the expansion rate of the Universe is well described by 
\mbox{$H(z) = H_0\sqrt{\Omega_\Lambda + \Omega_m(1+z)^3}$}, with the local expansion rate being $H_0 =70$~km~s$^{-1}$~Mpc$^{-1}$~\cite{Abdalla:2022yfr}, the dark-energy density $\Omega_\Lambda = 0.7$, and the matter density  $\Omega_m = 0.3$~\cite{Planck:2018vyg}.

We parametrize the CCSN  rate in terms of the initial mass function (IMF) and the star formation history $\dot{\rho}_*(z)$:
\begin{align}
    \mathcal{R}(z, M) = \frac{\text{IMF}(M)}{\int_{0.1 M_\odot}^{125 M_\odot} \text{d}M M\times \text{IMF}(M)}\dot{\rho}_*(z)\ ,
\end{align}
for collapsing stars with mass $\gtrsim 8 M_\odot$.
For what concerns the star formation history, we adopt the following piecewise parametrization~\cite{Horiuchi:2008jz}: 
\begin{align}
    \dot{\rho}_*(z) = \dot{\rho}_0 \Bigg[ &(1+z)^{\alpha\eta} \nonumber \\ & + \left(\frac{1+z}{B}\right)^{\beta\eta} + \left(\frac{1+z}{C}\right)^{\gamma\eta}\Bigg]^{1/\eta}\,,
    \label{eq:sfh}
\end{align}
with $\eta = -10$. The constants $B$ and $C$ are given by
\begin{align}
    &B = (1 + z_1)^{1-\alpha/\beta} \, , \\
    &C = (1 + z_1)^{(\beta - \alpha)/\gamma} (1 + z_2)^{1-\beta/\gamma} \,.
\end{align}
Given our poor knowledge of the CCSN rate, which constitutes the largest uncertainty plaguing the DSNB~\cite{Lunardini:2012ne,Horiuchi:2008jz}, we consider a  range of values for $\dot{\rho}_0$, $\alpha$, $\beta$, and $\gamma$ as summarized in Table~\ref{tab:sfh}, and we set $z_1=1$ and $z_2 = 4$.
\begin{table}[]
    \centering
    \caption{Numerical values of the parameters in the upper, fiducial, and lower analytical fits of the star formation rate from Ref.~\cite{Horiuchi:2008jz}, assuming the Salpeter initial mass function~\cite{Salpeter:1955it}.}
    \label{tab:sfh}
    \renewcommand{\arraystretch}{1.4}
    \begin{ruledtabular}
    \begin{tabular}{lcccc}
    Ranges& $\dot{\rho}_0$ & $\alpha$ & $\beta$ & $\gamma$\\ \colrule
    Upper & $0.0213$ &$3.6$  & $-0.1$ & $-2.5$ \\
    Fiducial & $0.0178$ & $3.4$  & $-0.3$  & $-3.5$ \\
    Lower & $0.0142$ & $3.2$ & $-0.5$ & $-4.5$ \\ 
    \end{tabular}
    \end{ruledtabular}
\end{table}
The analytical fits of $\dot{\rho}_*(z)$ depend on the assumed initial mass function~\cite{Horiuchi:2008jz}. We consider for the latter the  Salpeter initial mass function~\cite{Salpeter:1955it}: $\text{IMF} \propto (M/M_\odot)^{-2.35} $. Employing non-universal  initial mass functions  would affect  the CCSN rate  (i.e., $\dot{\rho}_0$) with a negligible impact on the DSNB~\cite{Hopkins:2006bw,Ziegler:2022ivq}. 

\paragraph*{Fraction of neutrino-driven black hole-forming collapses.} Another unknown in the DSNB modeling is the fraction of  non-rotating (or mildly-rotating) massive stars collapsing into BHs.
To take this into account, we allow for the  fraction of neutrino-driven BH-forming collapses ($f_{\nu\text{BH}}$) to vary, following Ref.~\cite{Moller:2018kpn}, and restrict the range to $f_{\nu\text{BH}} < 0.47$. This choice 
is motivated by the findings of Ref.~\cite{Kresse:2020nto}, which considers a suite of parameterized 1D  neutrino-driven engines for  single-star and helium-star progenitors  (i.e., $0.178 < f_{\nu\text{BH}}< 0.417$). 

\paragraph*{Neutrino flavor conversion.} The DSNB  is expected to be affected by neutrino flavor conversions~\cite{Lunardini:2012ne}. However, due to the large uncertainties in the modeling of neutrino quantum kinetics in the core of SNe~\cite{Tamborra:2020cul,Richers:2022zug,Mirizzi:2015eza}, we rely on a conservative approach and consider two extreme cases: (i) no flavor conversion, or (ii) full flavor conversion (i.e., $\nu_e$ or $\bar{\nu}_e$ emerge from the source as $\nu_x$ or $\bar\nu_x$, swapping completely their flavor). We stress that  none of these assumptions is realistic, yet they should serve as guidance to assess  the impact of flavor conversion uncertainties. 
Conversely, flavor conversion of neutrinos en route to Earth is not subject to large uncertainties~\cite{Dighe:1999bi}.
To model this, we consider the best-fit values of the neutrino mixing parameters for normal mass ordering from Ref.~\cite{deSalas:2020pgw}.

The dark gray band in the bottom panel of Fig.~\ref{fig:model-prediction} represents   the $\bar\nu_e$ component of our ``fiducial'' DSNB flux model, obtained for $0 < f_{\nu\text{BH}}  <  0.47$,  $0.0142 < \dot{\rho}_0 < 0.0213$, $3.2 < \alpha < 3.6$, $-0.5 < \beta < -0.1$, and $-4.5 < \gamma < -2.5$, as well as no flavor conversion. The light gray band of Fig.~\ref{fig:model-prediction} instead corresponds to the case of  full flavor conversion.
We assume the values for the fiducial parametrization of the CCSN  rate (cf.~Table~\ref{tab:sfh}) and $f_{\nu\text{BH}} = 0.3$, which is an average value according to Ref.~\cite{Kresse:2020nto}.
In order to make this plot, we consider each of the core collapse models from our suite as representative of the  population for a certain mass range (cf.~the top panel of Fig.~\ref{fig:imf_che}). The slope of the tail of the spectrum becomes less steep as  the   fraction of neutrino-driven BH-forming collapses increases, since the emission from this source class is more energetic. 

\subsection{Contribution to the diffuse neutrino background from  protomagnetars and spinars}
Assuming that our protomagnetar model A17 or our spinar model A26 are representative of the population of massive stellar collapses, the top panel of Fig.~\ref{fig:model-prediction} shows the corresponding $\bar\nu_e$ fluxes.~\footnote{We have verified that the neutrino spectral energy distributions of our suite of magnetorotational collapse models follow the parametrization of Refs.~\cite{Keil:2002in,Tamborra:2012ac}.} One can see that the $\bar\nu_e$ spectra from protomagnetars and spinars exhibit an energy tail with slope comparable to the one of the neutrino-driven BH-forming collapse model. However, for magnetorotational core collapses, the energy distributions  peak at $E \simeq 10$~MeV in the source frame.

When taking into account the contribution of protomagnetars and spinars, the total diffuse neutrino emission is
\begin{align}
    \Phi_\text{diffuse} = (1 -f_\text{MR})\Phi_{\nu\text{CCSN}} + f_\text{MR}\Phi_\text{MR}\ ,
    \label{eqn:total-flux}
\end{align}
where $\Phi_{\nu\text{CCSN}}$ is the diffuse flux from neutrino-driven CCSNe and neutrino-driven BH forming collapses (Eq.~\ref{eq:ccsn-dsnb}), while $\Phi_\text{MR}$ represents  the contribution from protomagnetars and spinars, and $f_\text{MR}$ is the fraction of magnetorotational collapses assumed to be constant over redshift.

\begin{figure}
    \centering
    \includegraphics[width=\linewidth]{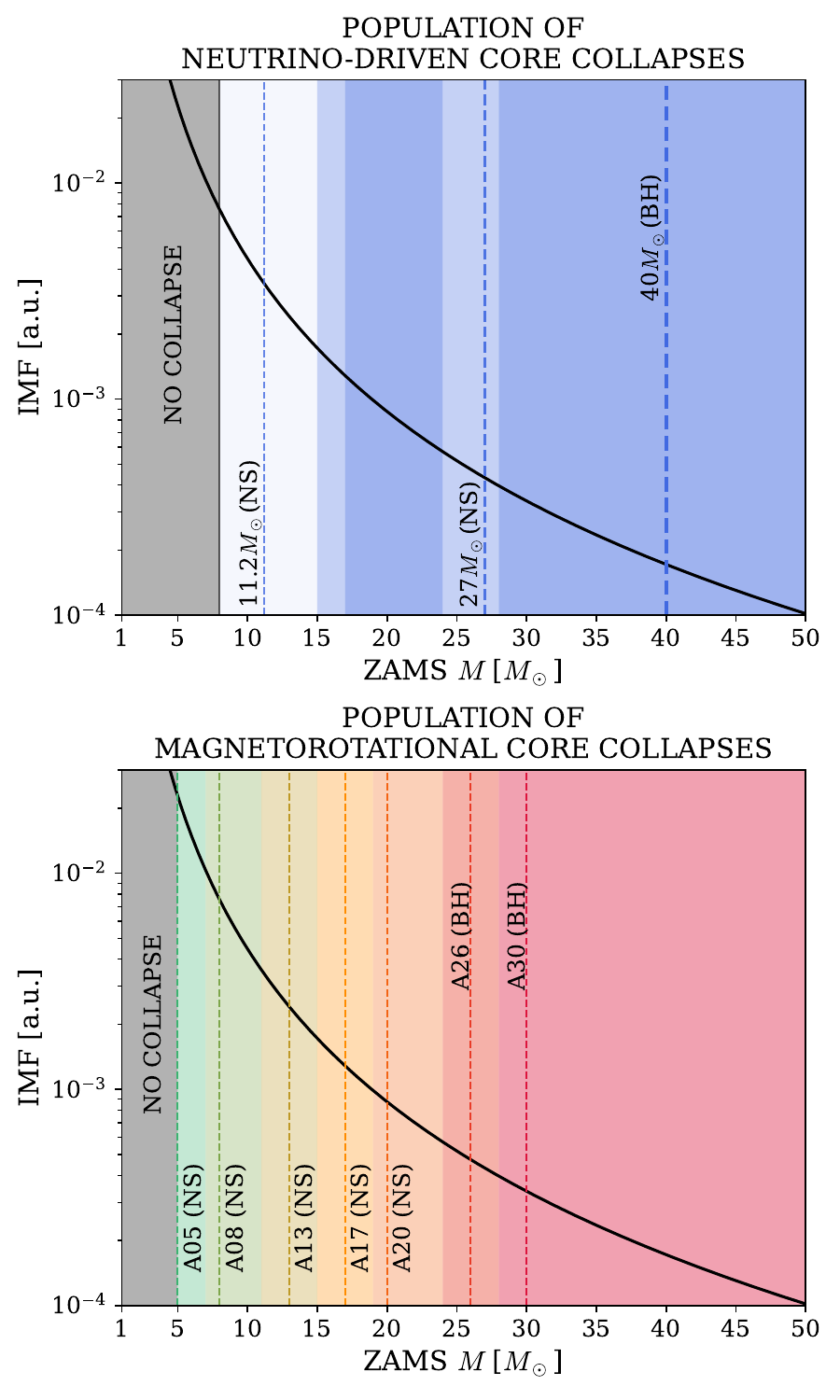}
    \caption{{\it Top panel:} Initial mass function for neutrino-driven CCSNe  and  BH forming collapses as a function of the ZAMS 
    progenitor mass. {\it Bottom panel:} Same as the top panel, but for  the magnetorotational core collapses. The neutrino-driven core collapse  and the protomagnetar/spinar  populations extend  up to $125 M_\odot$, although not explicitly shown in the plots. The vertical dashed lines highlight the ZAMS mass of the models in our suite. The color bands correspond to the range of masses for which we consider each model of our suite as representative. In parenthesis, we indicate if the remnant compact object is a neutron star (NS) or a BH, following Refs.~\cite{Mirizzi:2015eza,Garching} and the 3D models for the magnetorotational collapses adopted in this work (note that this setup fixes the fraction of spinars to $f_\text{BH-MR}\simeq 0.11$). 
    }
    \label{fig:imf_che}
\end{figure}

For the initial mass function  of magnetorotational stellar core collapses, we adopt the Salpeter one, but extending from $5 M_\odot$ to  $125 M_\odot$.
The lower minimum mass that we consider for magnetorotational collapses ($5 M_\odot$)  is motivated by the fact that the chemically homogeneous evolution of the models under consideration results in larger carbon-oxygen cores compared to those in non-rotating, standard core-collapse progenitors. These cores provide conditions conducive to the development of subsequent nuclear burning stages, even below the non-rotating threshold of approximately $8 M_\odot$. Consequently, chemically homogeneous  models in the mass range $[5 M_\odot, 8 M_\odot]$ may also undergo core collapse instead of evolving into carbon-oxygen white dwarfs.
The lower panel in Fig.~\ref{fig:imf_che} illustrates how the magnetorotational models introduced in Sec.~\ref{sec:models} are  employed to model the magnetorotational core collapse  population, where for each band in the initial mass function we have considered the correspondent protomagnetar or spinar model.  

We have also considered two extreme initial mass functions, $\text{IMF}_{\rm MR, 1}\propto (M/M_\odot)^{-2.7}$ and $\text{IMF}_\text{MR, 2}\propto (M/M_\odot)^{-1}$, following Refs.~\cite{Chabrier_2014} and \cite{Sibony_2022A&A...666A.199}, respectively. We find that the DSNB is negligibly affected by the choice of the IMF, in agreement with the findings of Ref.~\cite{Ziegler:2022ivq}; we, thus, only report our findings for the Salpeter mass function. 

The diffuse neutrino emission from protomagnetars and spinars is obtained analogously to Eq.~\ref{eq:ccsn-dsnb}.
It is worth noticing  that the prefactor $f_\text{MR}$ in Eq.~\ref{eqn:total-flux} accounts for the different normalization in the initial mass function of CCSNe  and the one of magnetars  resulting from the different lower mass limit ($8\,M_\odot$ or $5\,M_\odot$) adopted for each of the two cases.
In addition, in the computation of the contribution of neutrinos from magnetorotational collapses  to the DSNB, the uncertainties in the star formation history and flavor conversion  are taken into account as described in Sec.~\ref{subsec:dsnb-ccsn}.

The lower panel of Fig.~\ref{fig:model-prediction} shows the contribution of different fractions of magnetorotational collapses to the DSNB, one can see that the tail of the latter increases with the fraction of protomagnetars and spinars. It is worth noticing that  the  magnetorotational collapse models predict an emission of non-electron flavor neutrinos and antineutrinos with larger mean energies than the electron-flavor ones. Hence, uncertainties due to flavor conversion in the source can potentially  affect the  spectral shape of the related diffuse neutrino emission even more than in CCSNe.

\begin{figure}
    \centering
    \includegraphics[width=\linewidth]{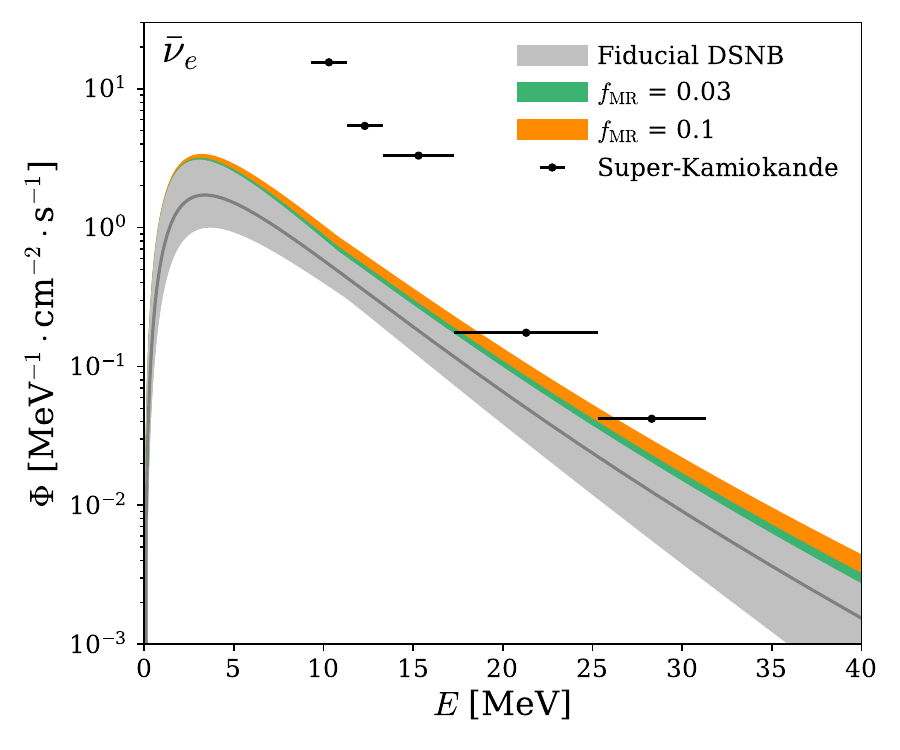}
    \caption{Electron-antineutrino DSNB  for $f_\text{MR}= 0.03$ (green) and $f_\text{MR} =0.1$ (orange), including the contribution from neutrino-driven CCSNe and  BH forming collapses and in the absence of flavor conversion. The gray band corresponds to the fiducial DSNB prediction (i.e., for $f_\text{MR} = 0.1$,  uncertainties on the star formation history as from Table~\ref{tab:sfh}, and $f_{\nu\text{BH}} < 0.47$). The solid gray line corresponds to the fiducial cosmic rate (cf.~Table~\ref{tab:sfh}) and $f_{\nu\text{BH}} =0.3$. The error bars represent the  model-independent upper limits from Super-Kamiokande on a flux of $\bar\nu_e$'s of astrophysical origin~\cite{SK-Gd:2024}. The neutrino emission from magnetorotational collapses enhances the high-energy tail of the DSNB spectrum, boosting its detection prospects.}
    \label{fig:diffuse-cases}
\end{figure}

Figure~\ref{fig:diffuse-cases} displays our $\bar\nu_e$ DSNB spectrum for $f_\text{MR} = 0.1$ and $f_\text{MR} = 0.03$. 
These fractions of magnetorotational collapses are chosen as representative. In fact, Ref.~\cite{2024ApJ...970..113N} finds that the fraction of fast rotating Be stars (CBe stars) is $(32.8\pm 3.4)\%$ for stars  with mass above $15$--$16 M_\odot$, and $(4.4 \pm 0.9)\%$ for  stars  with mass between $\sim 2$--$16 M_\odot$. Since an initially fast rotating star is   necessary for the formation of a protomagnetar,  the fraction of  fast-rotating CBe stars could be considered as an upper bound on the fraction of massive stars expected to  form protomagnetars.
Following Ref.~\cite{2024ApJ...970..113N}, we find that the IMF weighted average fraction of fast rotating stars in the  $[5 M_\odot, 26 M_\odot]$ mass range (where $26 M_\odot$ defines the minimum mass beyond which BH formation is expected in our 3D models) is $7.6\%$, which we approximate to $f_{\rm MR}=0.1$. 
The green and orange bands mostly overlap with the gray one, although one can clearly see that the contribution of magnetorotational collapses to the DSNB  enhances the high-energy tail of the spectrum. This effect is more visible as $f_\text{MR}$ increases (cf.~the bottom panel of Fig.~\ref{fig:model-prediction}).

\section{Constraining the fraction of magnetorotational collapses with the Super-Kamiokande-Gd data~\label{sec:current}}
Several neutrino telescopes carry out DSNB  searches, including KamLAND~\cite{KamLAND:2021gvi}, Borexino~\cite{Borexino:2019wln}, SNO~\cite{SNO:2004eru} and Super-Kamiokande~\cite{Super-Kamiokande:2020frs,Super-Kamiokande:2023xup}. Recently, the Super-Kamiokande Collaboration reported their most stringent  limits on a model-independent $\bar\nu_e$ flux of astrophysical origin using data from the phases VI and VII, with different levels of Gadolinium doping~\cite{SK-Gd:2024}.
We compare such limits to the  DSNB flux expected at Super-Kamiokande-Gadolinium computed as 
\begin{align}
    \bar{\Phi}(E_i, E_f) = \frac{\int_{E_i}^{E_f}\sigma_\text{IBD}(E) \Phi_\text{diffuse}(E)\text{d}E}{\int_{E_i}^{E_f}\sigma_\text{IBD}(E)\text{d}E}\ , 
    \label{eq:model-indep-flux}
\end{align}
where $\sigma_\text{IBD}$ is the cross-section for inverse beta decay~\cite{Strumia:2003zx,Ricciardi:2022pru}. For the sake of simplicity, we  assume an energy-independent detection efficiency and perfect energy resolution~\cite{Akhmedov:2022txm}. 

The large number of reactor antineutrinos below $10$~MeV and the atmospheric background above $\sim 30$~MeV constrain the neutrino detection window for the DSNB~\cite{Beacom:2010kk,Lunardini:2010ab}. The model-independent results reported by the Super-Kamiokande Collaboration account for these backgrounds and assume an energy-independent flux per energy bin. As a consequence, when comparing with our theoretical predictions, we unfold the energy-dependence of the detection cross section--see Eq.~\eqref{eq:model-indep-flux}. 

\begin{figure}
    \centering
    \includegraphics[width=\linewidth]{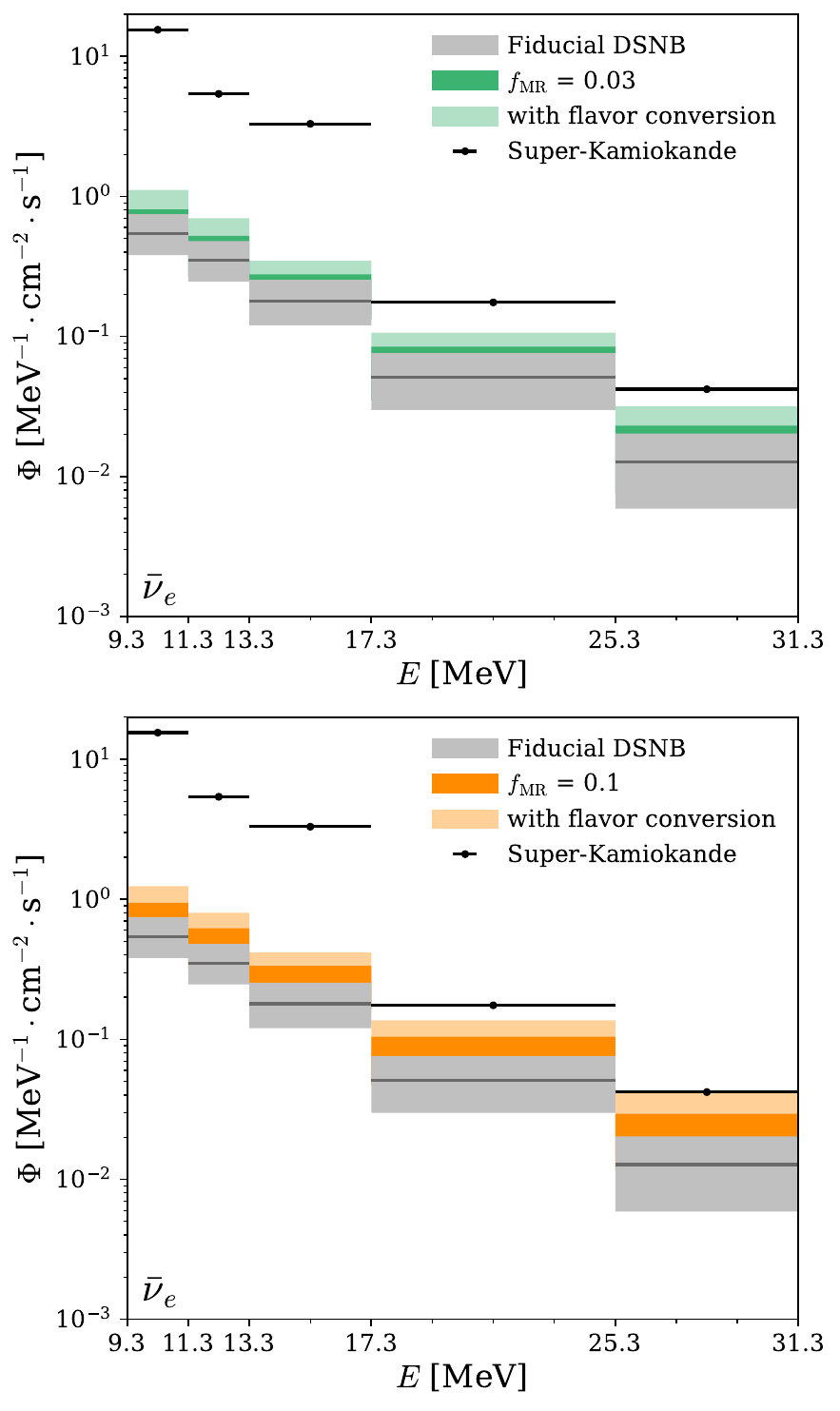}
    \caption{Electron-antineutrino DSNB flux binned in energy  for $f_\text{MR}=0.03$ (top panel) and $f_\text{MR}=0.1$ (bottom panel). The gray bands correspond to the fiducial DSNB model (i.e., obtained considering  uncertainties on the star formation history as from Table~\ref{tab:sfh}), $f_{\nu\text{BH}} < 0.47$, and no flavor conversion at the source. The dark shaded colored bands correspond to the DSNB obtained for  $f_\text{MR} =0.03$ and $f_\text{MR} =0.1$. Light shaded bands are obtained for  full flavor conversion at the source. The black dots and lines represent the  $\bar\nu_e$ model-independent limits obtained by  the Super-Kamiokande Collaboration~\cite{SK-Gd:2024}. Current Super-Kamiokande data do not allow to constrain $f_\text{MR} \lesssim 0.1$. }
    \label{fig:diffuse-obs}
\end{figure}

Figure~\ref{fig:diffuse-obs} compares the limits reported by the Super-Kamiokande Collaboration with our DSNB forecast. We consider the fiducial DSNB model ($f_\text{MR} = 0$, gray band), the DSNB computed for $f_\text{MR} = 0.03$ (top panel, dark green bands) and $f_\text{MR} = 0.1$ (bottom panel, dark orange bands) and also scenarios with full flavor conversion at the source (see Sec.~\ref{sec:dsnb}, light green and light orange bands, respectively). 
In all cases, our forecast for the DSNB signal falls  below the current Super-Kamiokande limits. 
However, if $f_\text{MR} \gtrsim 0.09$ and full flavor conversion would take place for the  highest local star formation rate, the Super-Kamiokande-Gadolinium upper limits would already be incompatible with the DSNB model. 
Note that, in order to allow for a direct quantitative comparison of our model to the data, the energy binning used in Fig.~\ref{fig:diffuse-obs} is analogous to the one adopted by the Super-Kamiokande-Gadolinium Collaboration~\cite{SK-Gd:2024}.

\section{Detection prospects and future sensitivity \label{sec:future}}
In this section, we discuss the impact of a non-zero fraction of magnetorotational core collapses in the near-future DSNB detection prospects  at Super-Kamiokande-Gadolinium~\cite{Simpson:2018snj} and JUNO~\cite{JUNO:2021vlw}. We also forecast the sensitivity of Hyper-Kamiokande-Gadolinium~\cite{Hyper-Kamiokande:2018ofw,Hyper-Kamiokande:2016srs} and  DUNE~\cite{DUNE:2020ypp} to  the fraction of magnetorotational collapses. The detector characteristics that we consider for these four neutrino telescopes are summarized in Table~\ref{tab:experiments}. This table also includes the number of signal events for $f_\text{MR}= 0.1$ and $f_\text{MR}= 0$ as well as the number of background events.

\subsection{Near term  detection prospects \label{sec:nearterm}}
The increase in  Gadolinium doping in  Super-Kamiokande has resulted in a larger event statistics  also for what concerns the DSNB~\cite{Marti:2019dof,Li:2022myd}. To illustrate the short-term potential of  Super-Kamiokande-Gadolinium, we  investigate the exposure that one would need to detect the DSNB and infer the fraction of magnetorotational collapses. 

\begin{figure}
    \centering
    \includegraphics[width=\linewidth]{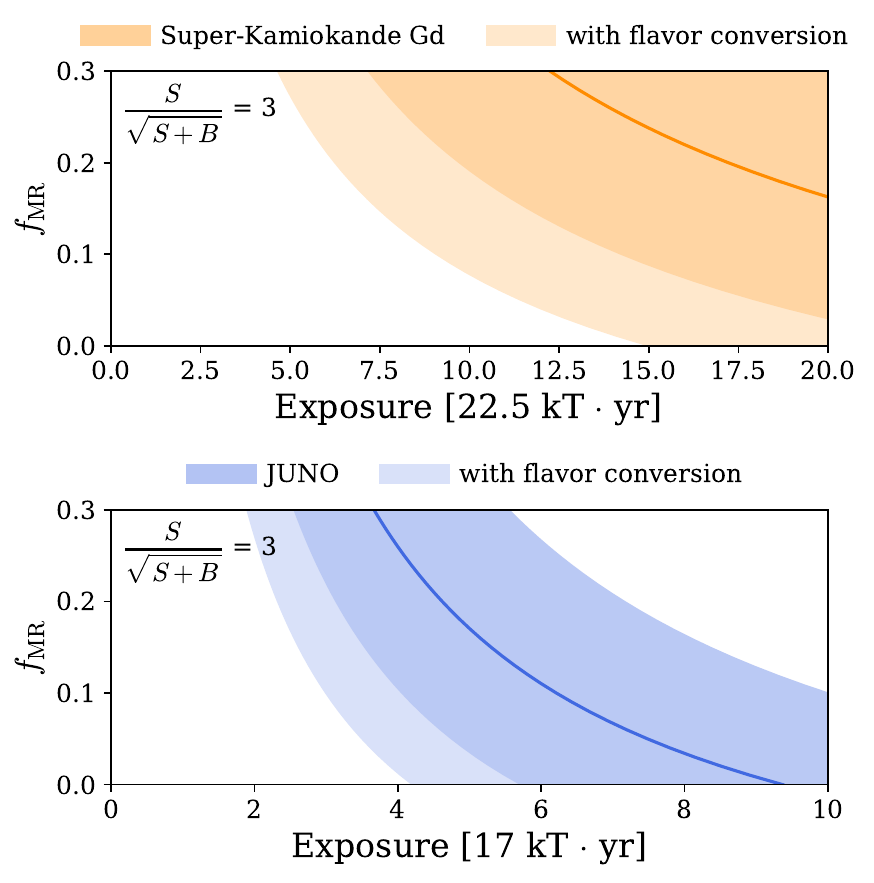}
    \caption{Fraction of magnetorotational collapses needed to reject the background-only hypothesis for  $S/\sqrt{S+B} = 3$ (cf.~main text for details) as a function of the exposure for Super-Kamiokande-Gadolinium (top panel) and JUNO (bottom panel). The shaded regions correspond to different sets of parameters for the star formation history (Table~\ref{tab:sfh}), considering no flavor conversion (darker shades) or full flavor conversion at the source (lighter shades).
    Under the assumption of full flavor conversion and for a $10$--$20\%$ fraction of magnetorotational core collapses, the detection of the DSNB could occur $2$--$4$~yr earlier than for  $f_\text{MR} = 0$.
    }
    \label{fig:detection-prospects}
\end{figure}

The expected number of signal events for a neutrino observatory is
\begin{align}
    S = N_t \tau \xi \int  \sigma_\text{IBD} (E)\Phi_\text{diffuse}(E) \text{d}E\,,
\end{align}
where $N_t$ is the number of targets for a given detector volume, $\tau$ is the  exposure time, and $\xi$ is the  energy-independent detection efficiency. The integral over the neutrino energy is carried out for the energy window of interest (see Table~\ref{tab:experiments})

For a fiducial volume of $22.5$~kT,  $55\%$ detection efficiency~\footnote{The detection efficiency is the result of a $75\%$ neutron-capture efficiency for a $0.03\%$ of Gd-loaded water and a $74\%$ selection efficiency~\cite{Super-Kamiokande:2023xup}.}, and the backgrounds reported for the Super-Kamiokande-VI and Super-Kamiokande-VII runs~\cite{SK-Gd:2024}, we compute the ratio between the signal events ($S$) and the Poissonian statistical fluctuations of  the total number of events, i.e.~signal plus background ($B$).  For our baseline analysis, the fraction of magnetorotational collapses is the only free parameter for a fixed star formation history (i.e., for a given set of values $\dot\rho_0$, $\alpha$, $\beta$ and $\gamma$; see Table~\ref{tab:sfh})--whereas the CCSN rate (see
Eq.~\ref{eq:sfh}) and  the fraction of neutrino-driven BH-forming collapses ($f_{\nu\text{BH}} = 0.3$) are assumed to be known. 

In the absence of systematic uncertainties, $S/\sqrt{S+B}=3$ corresponds to reject the background-only hypothesis with a significance of $3 \sigma$. Precisely, the top panel of Fig.~\ref{fig:detection-prospects} shows the region where $S/\sqrt{S+B}=3$, in the plane spanned by the exposure and the fraction of magnetorotational collapses.  The light-shaded band in the top panel of Fig.~\ref{fig:detection-prospects} corresponds to the case of full flavor conversion  at the source. 
Similarly, JUNO is expected to start  taking data in late 2024~\cite{Stock:2024tmd}. With a fiducial volume of $17$~kT, detection efficiency of $80\%$, and backgrounds as described in Ref.~\cite{JUNO:2022lpc}, its DSNB detection prospects  are comparable to the ones from Super-Kamiokande doped with Gadolinium, as visible from the bottom panel of Fig.~\ref{fig:detection-prospects}. For both Super-Kamiokande-Gadolinium and JUNO,  a  non-zero  $f_\text{MR}$ would result in a positive signal at neutrino observatories within less than a decade of data taking. In particular, a non-zero fraction of magnetorotational collapses could result in the  DSNB detection  within the next $6$--$8$~yr or earlier.

We stress that the  evolution of the sensitivity of Super-Kamiokande  depends on the  Gadolinium-enrichment phases and other experimental aspects such as background rejection and neutron tagging~\cite{Beacom:2003nk,Super-Kamiokande:2021the,Super-Kamiokande:2024kcb}. Changes in the experimental setup in JUNO may also modify the results presented above quantitatively. 

\begin{table*}[]
    \centering
    \caption{Fiducial volume, detection efficiency ($\xi$), and optimal energy window for JUNO~\cite{JUNO:2022lpc}, Super-Kamiokande-Gadolinium~(Super-K-Gd)~\cite{SK-Gd:2024}, DUNE~\cite{DUNE:2020ypp}, and Hyper-Kamiokande-Gadolinium~(Hyper-K-Gd)~\cite{Hyper-Kamiokande:2018ofw}. We also indicate the energy-integrated number of  DSNB  events for $f_\text{MR}=0$ and $f_\text{MR}=0.1$, and the number of background events per $100$~kT and year. }
    \renewcommand{\arraystretch}{1.4}
    \begin{ruledtabular}
    \begin{tabular}{lcccccc}
    Experiment& \begin{tabular}{@{}c@{}}Fiducial \\ volume \\ $[$kt$]$ \end{tabular}& \begin{tabular}{@{}c@{}}Detection \\ efficiency \\ $ $   \end{tabular}&\begin{tabular}{@{}c@{}}Energy \\window\\ $[$MeV$]$ \end{tabular}&  \begin{tabular}{@{}c@{}}DSNB events\\($f_\text{MR} = 0$)\\ $[(100$~ kT yr)$^{-1}]$\end{tabular} &  \begin{tabular}{@{}c@{}}DSNB events\\($f_\text{MR} = 0.1$) \\ $[(100$~kT yr)$^{-1}]$\end{tabular}  & \begin{tabular}{@{}c@{}}Background \\ events  \\ $[(100$~kT yr)$^{-1}]$\end{tabular}\\ \colrule
    JUNO & $17$ & $0.8$ & $9.4$--$31.4$ & $8.0$ & $11.0$ & $3.0$ \\
    Super-K-Gd & $22.5$ & $0.55$ & $9.4$-- $31.4$ & $5.2$ & $7.2$ & $36.1$ \\
    DUNE & $40$ & $1$ & $19.0$--$31.0$ & $1.3$ & $2.0$ & $0.6$ \\
    Hyper-K-Gd & $187$ & $0.55$ & $9.4$--$ 31.4$ & $5.2$ & $7.2$ & $18.0$ ($10.5$~\footnote{This value corresponds to the scenario in which  neutral current backgrounds induced by atmospheric neutrinos are  negligible.})
    \end{tabular}
    \end{ruledtabular}
    \label{tab:experiments}
\end{table*}

Importantly, the combination of data from different neutrino observatories--affected by different systematics and backgrounds--could further advance the detection of the DSNB and   hint at the existence of a population of magnetorotational collapses contributing to the DSNB. For instance, combining Super-Kamiokande-Gadolinium and JUNO would result in an increase of the statistical significance of an eventual DSNB detection, as evident from the expected number of events summarized in Table~\ref{tab:experiments}. 

\subsection{Long term detection prospects}
A better assessment on the contribution from protomagnetars and spinars to the DSNB is expected from the upcoming water Cherenkov neutrino telescope Hyper-Kamiokande~\cite{Hyper-Kamiokande:2018ofw}. 
We consider an experimental configuration with one tank of $187$~kT detection efficiency of $55\%$, resulting from adding Gd$_2$(SO$_4$)$_3\cdot$8H$_2$O, i.e for a 0.033\% of Gd-loaded water (cf.~Table~\ref{tab:experiments} for a summary on the expected number of events). Among the different backgrounds expected in Hyper-Kamiokande, neutral-current atmospheric background events are particularly relevant~\cite{Kunxian:2016joi,Ashida:2020erk}, and the possibility of reducing them to negligible level is under investigation~\cite{Maksimovic:2021dmz}.

\begin{figure}
    \centering
    \includegraphics[width=\linewidth]{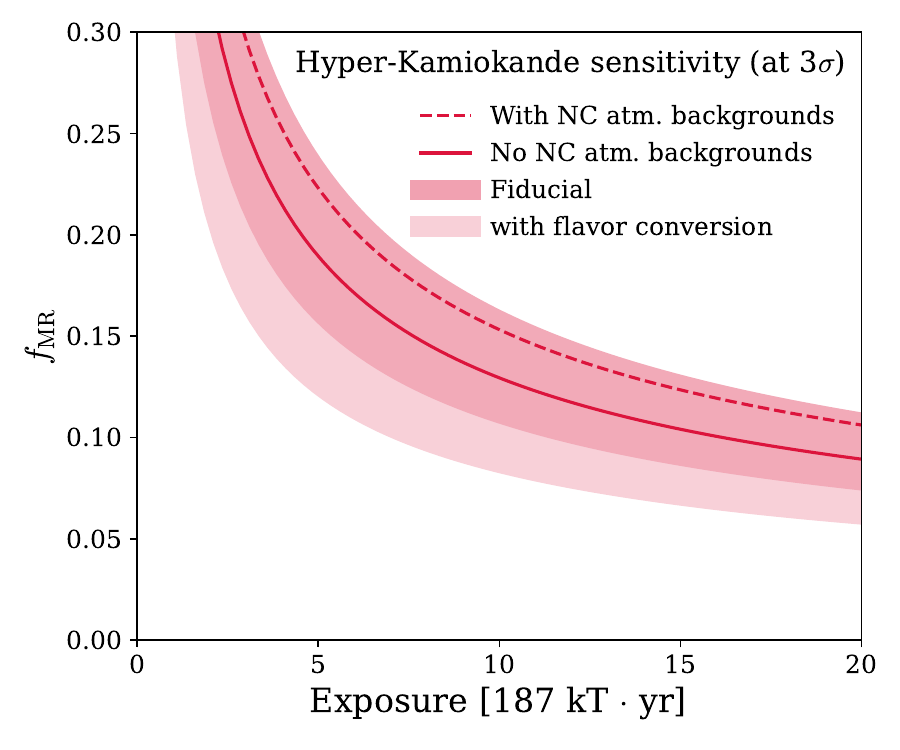}
    \caption{Hyper-Kamiokande sensitivity at $3 \sigma$ to the fraction of magnetorotational core collapses ($f_\text{MR}$)  as a function of the exposure. The solid line corresponds to negligible  neutral-current backgrounds induced by atmospheric neutrinos. The dashed line considers the fiducial backgrounds from Ref.~\cite{Martinez-Mirave:2024hfd}. The dark shaded band represents the impact of the uncertainties on the CCSN rate  and the fraction of neutrino-driven BH-forming collapses. After $20$~yr exposure, Hyper-Kamiokande-Gadolinium will be sensitive to $f_{\rm{MR}} \gtrsim 11\%$}.\label{fig:hk_senstivity}
\end{figure}

Figure~\ref{fig:hk_senstivity}  shows the sensitivity  to the fraction of magnetorotational collapses as a function of the exposure for the  DSNB model of Sec.~\ref{sec:dsnb}, with and without neutral current atmospheric backgrounds. This corresponds to the sensitivity to distinguish the DSNB prediction for a non-zero fraction of magnetorotational collapses from a DNSB signal with $f_\text{MR}=0$, i.e. we consider a ratio:
\begin{align}
    \frac{S(f_\text{MR}) - S(f_\text{MR}=0)}{\sqrt{S(f_\text{MR}) + B}}=3
    \, .
\end{align} 

One can see that a fraction of magnetorotational collapses smaller than $11\%$ ($16\%$) could be measured at  $3 \sigma$  after $20$~yr (10 yr) of data taking, assuming that the fraction of neutrino-driven BH-forming collapses is known and the neutral current atmospheric background is efficiently tagged. 

The upcoming liquid Argon neutrino experiment DUNE~\cite{DUNE:2020ypp}, being sensitive to the $\nu_e$ component, could also add positively to this picture, although with smaller statistics (cf.~Table~\ref{tab:experiments}). 
We estimate that a ratio $S/\sqrt{S+B}=3$ 
(i.e. to reject the background-only hypothesis at $3\sigma$ level) could only be achieved in $10$~yr at this detector, if $f_\text{MR} \gtrsim 0.23$--assuming $100\%$ detection efficiency and the backgrounds from Ref.~\cite{Martinez-Mirave:2024hfd}. Note that these sensitivity prospects rely on optimistic detector characteristics (e.g.~perfect detection efficiency and a detector configuration foreseeing  four tanks of $10$~kt each, whereas only the deployment of two tanks  is envisioned for DUNE Phase I~\cite{DUNE:2024wvj}).
The difference between the mean energies of electron  and non-electron flavor neutrinos is larger than the one for antineutrinos. This  could potentially disentangle some of the existing uncertainties in the picture, especially for what concerns the flavor conversion physics and the fraction of neutrino-driven BH-forming collapses~\cite{Moller:2018kpn}.
However, even under optimistic assumptions, the contribution of DUNE to this matter is likely to be marginal.

\section{Complementary electromagnetic constraints\label{sec:em}}
\begin{figure*}
    \centering
    \includegraphics[width=0.85\linewidth]{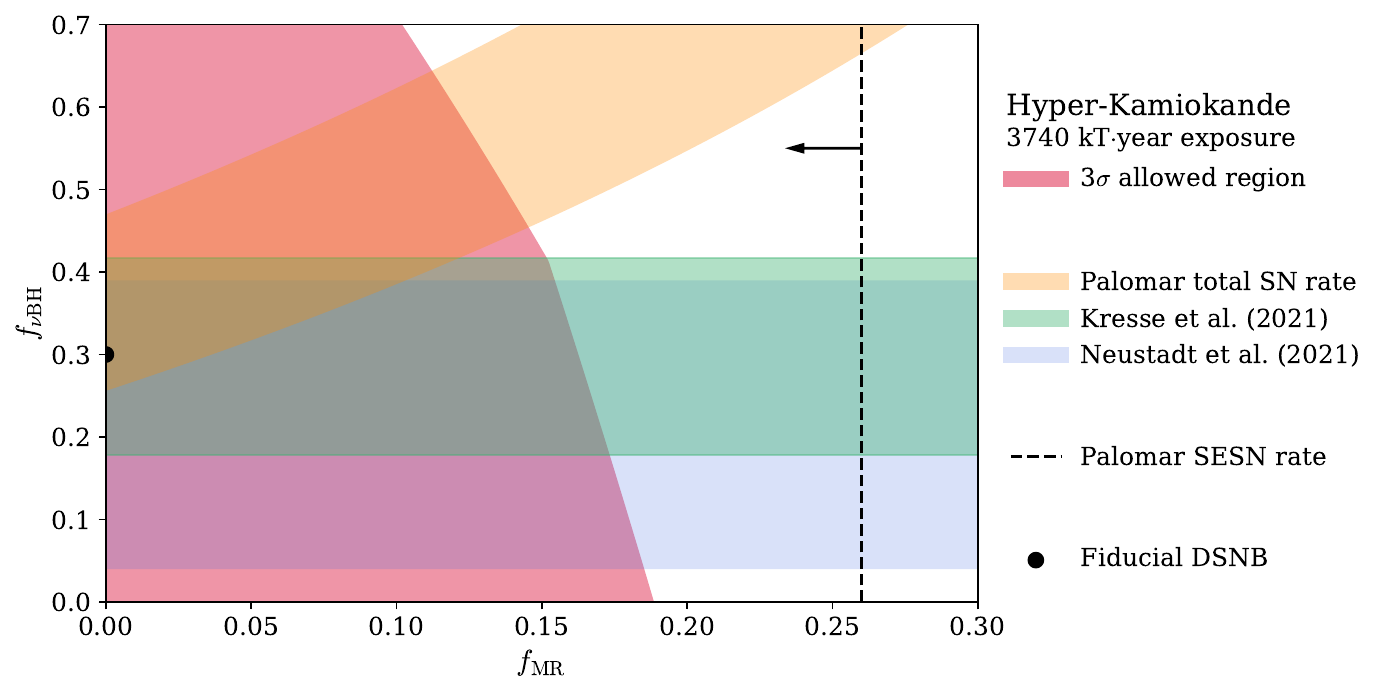}
    \caption{Present and future constraints from neutrino (DSNB) 
    electromagnetic data  and numerical simulations on the fraction of magnetorotational collapses ($f_\text{MR}$) and neutrino-driven BH-forming collapses ($f_{\nu\text{BH}}$). The red-shaded region corresponds to the projected sensitivity of Hyper-Kamiokande-Gadolinium after $20$~yr of data, taking into account  the  uncertainties on the CCSN rate. To guide the eye, the black dot marks our fiducial DSNB model. The orange band represents  the observed rate of successful explosions constrained by the Palomar facility~\cite{Frohmaier:2020pec}.
    The green-shaded region displays the interval of  $f_{\nu\text{BH}}$ extracted from simulations of the CCSN population~\cite{Kresse:2020nto}. Additional constraints from the search for disappearing luminous stars from the Large Binocular Telescope~\cite{Neustadt:2021a} are shown in blue.
    The vertical dashed line corresponds to the upper limits on the observed rate of stripped-envelope SNe (SESNe) at the Palomar transient facility~\cite{Frohmaier:2020pec}. 
    The neutrino and electromagnetic data and numerical simulations can potentially provide complementary constraints on $f_\text{MR}$ and $f_{\nu\text{BH}}$. 
   }
    \label{fig:multimessenger}
\end{figure*}

Assuming magnetorotational core collapses are distributed following the star formation rate, their fraction and the one of neutrino-driven BH-forming collapses are  degenerate in the DSNB signal. In both cases, the larger the population, the higher would be the high-energy tail of the DSNB spectrum. 
To illustrate this, we assume that the true fractions of neutrino-driven BH-forming collapses and magnetorotational collapses are $f_{\nu\text{BH}} =0.3$ and $f_\text{MR} = 0$ and compute the expected number of DSNB events after $20$~yr of data taking at Hyper-Kamiokande-Gadolinium, in the absence of flavor conversion.
Figure~\ref{fig:multimessenger} shows the region of the parameter space (i.e.~values of $f_\text{MR}$ and $f_{\nu\text{BH}}$) compatible with our hypothesis. To guide the eye, we also display the allowed  fraction of neutrino-driven BH-forming collapses from the numerical simulations of Ref.~\cite{Kresse:2020nto} as a green-shaded contour. Note that limits on BH formation could also be inferred from searches with  the Large Binocular Telescope on disappearing luminous stars~\cite{Neustadt:2021a};  we select from Ref.~\cite{Neustadt:2021a} the constraints obtained assuming only one failed SN: $f_{\nu\text{BH}} = 0.16^{+0.23}_{
-0.12}$, as not all of the luminous stars in their sample may be  neutrino-driven BH-forming collapses.
These limits are displayed in Fig.~\ref{fig:multimessenger} (shaded blue region).

While we have focused  on  neutrinos, combining DSNB and electromagnetic data could help to break the degeneracy in the  fraction of magnetorotational collapses plaguing the DSNB data.
The Palomar Transient Facility has measured the total rate of observed SNe--both CCSN  and stripped-envelope SNe--to be~\cite{Frohmaier:2020pec}:
\begin{align}
    \mathcal{R}^\text{obs}_\text{Palomar, total} = 9.10^{+1.56}_{-1.27} \times 10^{-5} \text{yr}^{-1} \text{Mpc}^{-3}\,,
\end{align}
for \mbox{$H_0 = 70$ km$\cdot$s$^{-1} $Mpc$^{-1}$} and at an average redshift of $\langle z\rangle =0.028$. The fraction of BH-forming collapses as well as  the one of protomagnetars and spinars could be inferred relying on
the Palomar data, 
as shown in Fig.~\ref{fig:multimessenger}. 

The Palomar Transient Facility has also determined the local rate of stripped-envelope SNe to be~\cite{Frohmaier:2020pec}:
\begin{align}
    \mathcal{R}^\text{obs}_\text{Palomar, SESN} = 2.41^{+0.81}_{-0.64} \times 10^{-5} \text{yr}^{-1} \text{Mpc}^{-3}\,.
\end{align}
One can then constrain the fraction of magnetorotational collapses by requiring that the observable rate of protomagnetars and spinars  predicted in this work is smaller than the rate of stripped SNe observed by the Palomar Transient Facility. This would provide an upper limit, since it would rely on the assumption that all stripped SNe are linked to our suite of magnetorotational collapses, i.e.,  $f_{\rm MR}< \mathcal{R}^\text{obs}_\text{Palomar, SESN}/ \mathcal{R}^\text{obs}_\text{Palomar, total}\sim 0.26$.

\section{Conclusions  and outlook\label{sec:conclusions}}
As the detection of the DSNB approaches, it is timely to address whether the diffuse neutrino emission from magnetorotational core collapses, leading to the formation of protomagnetars or spinars, could contribute to the DSNB. 
We rely on a suite of seven 3D special relativistic neutrino-magnetohydrodynamic simulations with different ZAMS mass, inspired by the 2D equivalent models presented in Ref.~\cite{Obergaulinger:2021omt}, to investigate the related neutrino properties and the associated contribution to the DSNB. Our models   predict large neutrino luminosities $\gtrsim 10^{52}$ erg$\cdot$ s$^{-1}$ for several seconds and large  mean energies for non-electron neutrinos and antineutrinos ($\langle \varepsilon_{\nu_x} \rangle \gtrsim 20$~MeV). Such neutrino properties suggest that magnetorotational core collapses could provide a non-negligible contribution to the high-energy DSNB tail (above $5$~MeV). In this work, the evolution of the magnetorotational core collapses forming a protomagnetar is computed up to several seconds. A longer simulation would be desiderable to determine the evolution of the neutrino emission properties and  assess the impact of the late-time emission on the diffuse emission and  the potential numerical influence on the gradual rise of the mean energies at late times. The approach here adopted ensures that our results are conservative and do not overpredict the contribution of protomagnetars to the diffuse flux from MR core collapses.

We show that current data from Super-Kamiokande-Gadolinium are compatible with a fraction of magnetorotational collapses smaller than $\sim 9\%$ (assumed to be constant as a function of the redshift, for the sake of simplicity) after accounting for  uncertainties in the CCSN rate, the fraction of neutrino-driven BH-forming collapses, and  flavor conversion physics at the source. 
Due to the enhancement of the high-energy tail of the DSNB spectrum, a non-zero fraction of magnetorotational collapses could also lead to an earlier detection of the $\bar\nu_e$ component of the DSNB flux. In particular, in the near future, Super-Kamiokande-Gadolinium and JUNO could  reject the background-only hypothesis with an exposure  of $6$--$8$~yr for fractions of magnetorotational collapses above $10$--$20\%$, i.e.~$2$--$4$~yr earlier than for $f_\text{MR}=0$. 

Given its larger size, the upcoming  Hyper-Kamiokande-Gadolinium neutrino telescope could measure a fraction of magnetorotational collapses larger than $13\%$ at $3\sigma$ after $20$~yr of data-taking, if meanwhile we can infer  the fraction of (neutrino-driven) BH-forming collapses and pin down the uncertainties on the CCSN rate through electromagnetic observations. This measurement would further benefit from improved background reduction and detection efficiency.

The observational signatures of  magnetorotational collapses  in the DSNB signal are  entangled with the ones linked to neutrino-driven BH-forming collapses. However, complementary electromagnetic probes, e.g., searches for disappearing luminous stars~\cite{Gerke:2015a,Adams:2017a,Neustadt:2021a}, could help to break such degeneracies. 
Moreover, previous work~\cite{Aguilera-Dena:2018ork,Obergaulinger:2021omt} suggests that a subset of our suite of magnetorotational models could harbor progenitors of long gamma-ray bursts (LGRBs). 
Assuming that  the magnetorotational progenitors with ZAMS mass $M \gtrsim 15 M_\odot$ lead to GRB jets~\cite{Aguilera-Dena:2018ork} and that magnetorotational core collapses are the only sources of LGRBs, one could 
compare the rate of LGRBs with our findings to break the DSNB  degeneracies. Similarly, the connection between these magnetorotational collapses and superluminous SNe~\cite{Aguilera-Dena:2018ork} could also be explored.
Such  multi-messenger approach  would greatly benefit from reduced uncertainties on  the star formation history~\cite{Lien:2010yb} as well as progress in  stellar evolution models \cite{Paxton_2018ApJS..234...34, Ekstrom_2012A&A...537A.146,Heger_2003ApJ...591..288,Griffiths_2024arXiv240803368}, and our understanding of angular momentum transport in massive stars~\cite{Muller:2024slv, Griffiths_2022A&A...665A.147, Ma:2019cpr,Heger:2004qp}.

With this work, we set a roadmap for an improved comprehension of the sources that can contribute  to the DSNB, particularly protomagnetars and spinars. While existing data are still plagued by astrophysical uncertainties, the combination of upcoming neutrino and electromagnetic measurements will be crucial to learn more about the properties of the population of magnetorotational collapses in our Universe.  

\begin{acknowledgments}
In Copenhagen, this project has received support from the Villum Foundation 
(Project No.~13164), the Danmarks Frie Forskningsfond (Project No.~8049-00038B), the  Carlsberg Foundation (CF18-0183) and the Deutsche Forschungsgemeinschaft through Sonderforschungbereich SFB 1258 ``Neutrinos and Dark Matter in Astro- and Particle Physics'' (NDM). Part of this work was  performed  at Aspen Center for Physics, which is supported by National Science Foundation grant PHY-2210452. 
In Valencia, this work was supported  from the grant PID2021-127495NB-I00, funded by MCIN/AEI/10.13039/501100011033 and by the European Union under ``NextGenerationEU'' as well as ``ESF: Investing in your future'',   the Astrophysics and High Energy Physics program of the Generalitat Valenciana ASFAE/2022/026 funded by MCIN and the European Union ``NextGenerationEU'' (PRTR-C17.I1), and the Prometeo excellence program grant CIPROM/2022/13 funded by the Generalitat Valenciana. MO acknowledges support from the Spanish Ministry of Science via the Ram\'on y Cajal programme (RYC2018-024938-I). The Tycho supercomputer hosted at the SCIENCE HPC Center at the University of Copenhagen and computing time granted via the Red Espa\~{n}ola de Supercomputaci\'on (grants AECT-2022-3-0016, AECT-2023-1-0009, AECT-2023-3-0019, AECT-2024-1-0013) at the supercomputers MareNostrum-4 and MareNostrum-5 of the Barcelona Supercomputing Centre as well as on the hosts Tirant (grants AECT-2022-1-0003, AECT-2023-1-0001) and Lluisvives of the Servei d'Inform\`atica of the Universitat de Val\`encia were used for supporting the numerical simulations presented in this work. 
\end{acknowledgments}

\bibliography{bibliography}

@ARTICLE{Griffiths_2022A&A...665A.147,
       author = {{Griffiths}, Adam and {Eggenberger}, Patrick and {Meynet}, Georges and {Moyano}, Facundo and {Aloy}, Miguel {\'A}.},
        title = "{The magneto-rotational instability in massive stars}",
      journal = {\aap},
         year = 2022,
        month = sep,
       volume = {665},
          eid = {A147},
        pages = {A147},
          doi = {10.1051/0004-6361/202243599},
archivePrefix = {arXiv},
       eprint = {2204.00016},
 primaryClass = {astro-ph.SR},
       adsurl = {https://ui.adsabs.harvard.edu/abs/2022A&A...665A.147G}
}

@INPROCEEDINGS{1971swng.conf..485M,
       author = {{Morrison}, P. and {Cavaliere}, A.},
        title = "{Spinars - A Progress Report}",
    booktitle = {Study Week on Nuclei of Galaxies},
         year = 1971,
       editor = {{O'Connell}, D.~J.~K.},
        month = jan,
        pages = {485},
       adsurl = {https://ui.adsabs.harvard.edu/abs/1971swng.conf..485M}
}

@ARTICLE{1983A&A...127L...1L,
       author = {{Lipunov}, V.~M.},
        title = "{Detection of magnetomultipole radiation from neutron stars}",
      journal = {\aap},
         year = 1983,
        month = oct,
       volume = {127},
       number = {1},
        pages = {L1},
       adsurl = {https://ui.adsabs.harvard.edu/abs/1983A&A...127L...1L}
}

@BOOK{1987ans..book.....L,
       author = {{Lipunov}, Vladimir Mikhailovich},
        title = "{Astrophysics of neutron stars}",
         year = 1987,
  PUBLISHER = "{New York: Springer}",
       adsurl = {https://ui.adsabs.harvard.edu/abs/1987ans..book.....L}
}

@ARTICLE{1998A&A...329L..29L,
       author = {{Lipunova}, G.~V. and {Lipunov}, V.~M.},
        title = "{Formation of a gravitationally bound object after binary neutron star merging and GRB phenomena}",
      journal = {\aap},
         year = 1998,
        month = jan,
       volume = {329},
        pages = {L29-L32},
          doi = {10.48550/arXiv.astro-ph/9704057},
archivePrefix = {arXiv},
       eprint = {astro-ph/9704057},
 primaryClass = {astro-ph},
       adsurl = {https://ui.adsabs.harvard.edu/abs/1998A&A...329L..29L}
}

@ARTICLE{Heger_2003ApJ...591..288,
       author = {{Heger}, A. and {Fryer}, C.~L. and {Woosley}, S.~E. and {Langer}, N. and {Hartmann}, D.~H.},
        title = "{How Massive Single Stars End Their Life}",
      journal = {Astrophysical Journal},
         year = 2003,
        month = jul,
       volume = {591},
       number = {1},
        pages = {288-300},
          doi = {10.1086/375341},
archivePrefix = {arXiv},
       eprint = {astro-ph/0212469},
 primaryClass = {astro-ph},
       adsurl = {https://ui.adsabs.harvard.edu/abs/2003ApJ...591..288H}
}

@ARTICLE{Griffiths_2024arXiv240803368,
       author = {{Griffiths}, Adam and {Aloy}, Miguel-{\'A} and {Hirschi}, Raphael and {Reichert}, Moritz and {Obergaulinger}, Matrin and {Whitehead}, Emily E. and {Martinet}, S{\'e}bastien and {Esktr{\"o}m}, Sylvia and {Meynet}, Georges},
        title = "{Evolving massive stars to core collapse with GENEC: Extension of equation of state, opacities and effective nuclear network}",
      journal = {arXiv e-prints},
         year = 2024,
        month = aug,
          eid = {arXiv:2408.03368},
        pages = {arXiv:2408.03368},
          doi = {10.48550/arXiv.2408.03368},
archivePrefix = {arXiv},
       eprint = {2408.03368},
 primaryClass = {astro-ph.SR},
       adsurl = {https://ui.adsabs.harvard.edu/abs/2024arXiv240803368G}
}

@ARTICLE{Ekstrom_2012A&A...537A.146,
       author = {{Ekstr{\"o}m}, S. and {Georgy}, C. and {Eggenberger}, P. and {Meynet}, G. and {Mowlavi}, N. and {Wyttenbach}, A. and {Granada}, A. and {Decressin}, T. and {Hirschi}, R. and {Frischknecht}, U. and {Charbonnel}, C. and {Maeder}, A.},
        title = "{Grids of stellar models with rotation. I. Models from 0.8 to 120 M$_{{\ensuremath{\odot}}}$ at solar metallicity (Z = 0.014)}",
      journal = {Astronomy \& Astrophysics,},
         year = 2012,
        month = jan,
       volume = {537},
          eid = {A146},
        pages = {A146},
          doi = {10.1051/0004-6361/201117751},
archivePrefix = {arXiv},
       eprint = {1110.5049},
 primaryClass = {astro-ph.SR},
       adsurl = {https://ui.adsabs.harvard.edu/abs/2012A&A...537A.146E}
}

@ARTICLE{Paxton_2018ApJS..234...34,
       author = {{Paxton}, Bill and {Schwab}, Josiah and {Bauer}, Evan B. and {Bildsten}, Lars and {Blinnikov}, Sergei and {Duffell}, Paul and {Farmer}, R. and {Goldberg}, Jared A. and {Marchant}, Pablo and {Sorokina}, Elena and {Thoul}, Anne and {Townsend}, Richard H.~D. and {Timmes}, F.~X.},
        title = "{Modules for Experiments in Stellar Astrophysics (MESA): Convective Boundaries, Element Diffusion, and Massive Star Explosions}",
      journal = {Astrophysical Journal, Supplement},
         year = 2018,
        month = feb,
       volume = {234},
       number = {2},
          eid = {34},
        pages = {34},
          doi = {10.3847/1538-4365/aaa5a8},
archivePrefix = {arXiv},
       eprint = {1710.08424},
 primaryClass = {astro-ph.SR},
       adsurl = {https://ui.adsabs.harvard.edu/abs/2018ApJS..234...34P}
}

@ARTICLE{Sibony_2022A&A...666A.199,
       author = {{Sibony}, Y. and {Liu}, B. and {Simmonds}, C. and {Meynet}, G. and {Bromm}, V.},
        title = "{Impact of Population III homogeneous stellar evolution on early cosmic reionisation}",
      journal = {\aap},
         year = 2022,
        month = oct,
       volume = {666},
          eid = {A199},
        pages = {A199},
          doi = {10.1051/0004-6361/202244146},
archivePrefix = {arXiv},
       eprint = {2205.15125},
 primaryClass = {astro-ph.SR},
       adsurl = {https://ui.adsabs.harvard.edu/abs/2022A&A...666A.199S}
}

@ARTICLE{Metzger_2011MNRAS.413.2031,
       author = {{Metzger}, B.~D. and {Giannios}, D. and {Thompson}, T.~A. and {Bucciantini}, N. and {Quataert}, E.},
        title = "{The protomagnetar model for gamma-ray bursts}",
      journal = {\mnras},
         year = 2011,
        month = may,
       volume = {413},
       number = {3},
        pages = {2031-2056},
          doi = {10.1111/j.1365-2966.2011.18280.x},
archivePrefix = {arXiv},
       eprint = {1012.0001},
 primaryClass = {astro-ph.HE},
       adsurl = {https://ui.adsabs.harvard.edu/abs/2011MNRAS.413.2031M}
}

@ARTICLE{2024ApJ...970..113N,
       author = {{Navarete}, Felipe and {Ticiani dos Santos}, Pedro and {Carciofi}, Alex Cavali{\'e}ri and {Figueiredo}, Andr{\'e} Luiz},
        title = "{On the Origin of Fast-rotating Stars. I. Photometric Calibration and Results of AO-assisted BVRI+H{\ensuremath{\alpha}} Imaging of NGC 330 with SAMI/SOAR}",
      journal = {\apj},
         year = 2024,
        month = aug,
       volume = {970},
       number = {2},
          eid = {113},
        pages = {113},
          doi = {10.3847/1538-4357/ad500f},
archivePrefix = {arXiv},
       eprint = {2405.12429},
 primaryClass = {astro-ph.GA},
       adsurl = {https://ui.adsabs.harvard.edu/abs/2024ApJ...970..113N}
}

@ARTICLE{Metzger_2018ApJ...857...95,
       author = {{Metzger}, Brian D. and {Beniamini}, Paz and {Giannios}, Dimitrios},
        title = "{Effects of Fallback Accretion on Protomagnetar Outflows in Gamma-Ray Bursts and Superluminous Supernovae}",
      journal = {\apj},
         year = 2018,
        month = apr,
       volume = {857},
       number = {2},
          eid = {95},
        pages = {95},
          doi = {10.3847/1538-4357/aab70c},
archivePrefix = {arXiv},
       eprint = {1802.07750},
 primaryClass = {astro-ph.HE},
       adsurl = {https://ui.adsabs.harvard.edu/abs/2018ApJ...857...95M}
}

@article{Obergaulinger:2021omt,
author = {{Obergaulinger}, M. and {Aloy}, M. {\'A}.},
        title = "{Magnetorotational core collapse of possible gamma-ray burst progenitors - IV. A wider range of progenitors}",
      journal = {\mnras},
         year = 2022,
        month = may,
       volume = {512},
       number = {2},
        pages = {2489-2507},
          doi = {10.1093/mnras/stac613},
archivePrefix = {arXiv},
       eprint = {2108.13864},
 primaryClass = {astro-ph.HE},
       adsurl = {https://ui.adsabs.harvard.edu/abs/2022MNRAS.512.2489O}
}

@inproceedings{Bisnovatyi-Kogan:1982oyy,
    author = "Bisnovatyi-Kogan, G. S. and Seidov, Z. F.",
    title = "{SUPERNOVAE, NEUTRINO REST MASS, AND THE MIDDLE ENERGY NEUTRINO BACKGROUND IN THE UNIVERSE}",
    booktitle = "{11th Texas Symposium on Relativistic Astrophysics}",
    year = "1982"
}

@ARTICLE{1984SvA....28...30D,
       author = {{Domogatskii}, G.~V.},
        title = "{The Isotropic Electron Antineutrino Flux - a Clue to the Rate of Stellar Gravitational Collapse in the Universe}",
      journal = {Soviet Astronomy},
         year = 1984,
        month = feb,
       volume = {28},
        pages = {30},
       adsurl = {https://ui.adsabs.harvard.edu/abs/1984SvA....28...30D}
}

@article{Lien:2010yb,
    author = "Lien, Amy and Fields, Brian D. and Beacom, John F.",
    title = "{Synoptic Sky Surveys and the Diffuse Supernova Neutrino Background: Removing Astrophysical Uncertainties and Revealing Invisible Supernovae}",
    eprint = "1001.3678",
    archivePrefix = "arXiv",
    primaryClass = "astro-ph.CO",
    doi = "10.1103/PhysRevD.81.083001",
    journal = "Phys. Rev. D",
    volume = "81",
    pages = "083001",
    year = "2010"
}

@article{Muller:2024slv,
    author = {M\"uller, Bernhard},
    title = "{Supernova Simulations}",
    eprint = "2403.18952",
    archivePrefix = "arXiv",
    primaryClass = "astro-ph.HE",
    month = "3",
    journal="",
    year = "2024"
}

@article{Ma:2019cpr,
    author = "Ma, Linhao and Fuller, Jim",
    title = "{Angular momentum transport in massive stars and natal neutron star rotation rates}",
    eprint = "1907.03713",
    archivePrefix = "arXiv",
    primaryClass = "astro-ph.SR",
    doi = "10.1093/mnras/stz2009",
    journal = "Mon. Not. Roy. Astron. Soc.",
    volume = "488",
    number = "3",
    pages = "4338--4355",
    year = "2019"
}

@article{Heger:2004qp,
    author = "Heger, Alexander and Woosley, S. E. and Spruit, H. C.",
    title = "{Presupernova evolution of differentially rotating massive stars including magnetic fields}",
    eprint = "astro-ph/0409422",
    archivePrefix = "arXiv",
    reportNumber = "LA-UR-04-6161",
    doi = "10.1086/429868",
    journal = "Astrophys. J.",
    volume = "626",
    pages = "350--363",
    year = "2005"
}

@article{Beacom:2003nk,
    author = "Beacom, John F. and Vagins, Mark R.",
    title = "{GADZOOKS! Anti-neutrino spectroscopy with large water Cherenkov detectors}",
    eprint = "hep-ph/0309300",
    archivePrefix = "arXiv",
    reportNumber = "FERMILAB-PUB-03-249-A",
    doi = "10.1103/PhysRevLett.93.171101",
    journal = "Phys. Rev. Lett.",
    volume = "93",
    pages = "171101",
    year = "2004"
}

@article{Super-Kamiokande:2023xup,
    author = "Harada, M. and others",
    collaboration = "Super-Kamiokande",
    title = "{Search for Astrophysical Electron Antineutrinos in Super-Kamiokande with 0.01\% Gadolinium-loaded Water}",
    eprint = "2305.05135",
    archivePrefix = "arXiv",
    primaryClass = "astro-ph.HE",
    doi = "10.3847/2041-8213/acdc9e",
    journal = "Astrophys. J. Lett.",
    volume = "951",
    number = "2",
    pages = "L27",
    year = "2023"
}

@article{Beacom:2010kk,
    author = "Beacom, John F.",
    title = "{The Diffuse Supernova Neutrino Background}",
    eprint = "1004.3311",
    archivePrefix = "arXiv",
    primaryClass = "astro-ph.HE",
    doi = "10.1146/annurev.nucl.010909.083331",
    journal = "Ann. Rev. Nucl. Part. Sci.",
    volume = "60",
    pages = "439--462",
    year = "2010"
}

@article{Steiner:2012rk,
    author = "Steiner, Andrew W. and Hempel, Matthias and Fischer, Tobias",
    title = "{Core-collapse supernova equations of state based on neutron star observations}",
    eprint = "1207.2184",
    archivePrefix = "arXiv",
    primaryClass = "astro-ph.SR",
    reportNumber = "INT-PUB-12-033",
    doi = "10.1088/0004-637X/774/1/17",
    journal = "Astrophys. J.",
    volume = "774",
    pages = "17",
    year = "2013"
}

@article{Tamborra:2020cul,
    author = "Tamborra, Irene and Shalgar, Shashank",
    title = "{New Developments in Flavor Evolution of a Dense Neutrino Gas}",
    eprint = "2011.01948",
    archivePrefix = "arXiv",
    primaryClass = "astro-ph.HE",
    doi = "10.1146/annurev-nucl-102920-050505",
    journal = "Ann. Rev. Nucl. Part. Sci.",
    volume = "71",
    pages = "165--188",
    year = "2021"
}

@inbook{Richers:2022zug,
    author = "Richers, Sherwood and Sen, Manibrata",
    editor = "Tanihata, Isao and Toki, Hiroshi and Kajino, Toshitaka",
    title = "{Fast Flavor Transformations}",
    booktitle = "{Handbook of Nuclear Physics}",
    eprint = "2207.03561",
    archivePrefix = "arXiv",
    primaryClass = "astro-ph.HE",
    doi = "10.1007/978-981-15-8818-1_125-1",
    pages = "1--17",
    year = "2022"
}

@article{Dighe:1999bi,
    author = "Dighe, Amol S. and Smirnov, Alexei {Yu}.",
    title = "{Identifying the neutrino mass spectrum from the neutrino burst from a supernova}",
    eprint = "hep-ph/9907423",
    archivePrefix = "arXiv",
    reportNumber = "IC-99-83",
    doi = "10.1103/PhysRevD.62.033007",
    journal = "Phys. Rev. D",
    volume = "62",
    pages = "033007",
    year = "2000"
}

@article{Keil:2002in,
    author = "Keil, Mathias Th. and Raffelt, Georg G. and Janka, Hans-Thomas",
    title = "{Monte Carlo study of supernova neutrino spectra formation}",
    eprint = "astro-ph/0208035",
    archivePrefix = "arXiv",
    doi = "10.1086/375130",
    journal = "Astrophys. J.",
    volume = "590",
    pages = "971--991",
    year = "2003"
}

@article{Tamborra:2012ac,
    author = "Tamborra, Irene and Mueller, Bernhard and Huedepohl, Lorenz and Janka, Hans-Thomas and Raffelt, Georg G.",
    title = "{High-resolution supernova neutrino spectra represented by a simple fit}",
    eprint = "1211.3920",
    archivePrefix = "arXiv",
    primaryClass = "astro-ph.SR",
    doi = "10.1103/PhysRevD.86.125031",
    journal = "Phys. Rev. D",
    volume = "86",
    pages = "125031",
    year = "2012"
}

@article{Lunardini:2010ab,
    author = "Lunardini, Cecilia",
    title = "{Diffuse supernova neutrinos at underground laboratories}",
    eprint = "1007.3252",
    archivePrefix = "arXiv",
    primaryClass = "astro-ph.CO",
    doi = "10.1016/j.astropartphys.2016.02.005",
    journal = "Astropart. Phys.",
    volume = "79",
    pages = "49--77",
    year = "2016"
}

@article{Mirizzi:2015eza,
    author = "Mirizzi, Alessandro and Tamborra, Irene and Janka, Hans-Thomas and Saviano, Ninetta and Scholberg, Kate and Bollig, Robert and Hudepohl, Lorenz and Chakraborty, Sovan",
    title = "{Supernova Neutrinos: Production, Oscillations and Detection}",
    eprint = "1508.00785",
    archivePrefix = "arXiv",
    primaryClass = "astro-ph.HE",
    doi = "10.1393/ncr/i2016-10120-8",
    journal = "Riv. Nuovo Cim.",
    volume = "39",
    number = "1-2",
    pages = "1--112",
    year = "2016"
}

@article{Ando:2023fcc,
author = {{Ando}, Shin'Ichiro and {Ekanger}, Nick and {Horiuchi}, Shunsaku and {Koshio}, Yusuke},
        title = "{Diffuse neutrino background from past core collapse supernovae}",
  eprint = {2306.16076},         
archivePrefix = {arXiv},
 primaryClass = "astro-ph.HE",
 doi = {10.2183/pjab.99.026},
      journal = "{Proceedings of the Japan Academy, Series B}",
  volume = {99},
  pages = {460-479},
   year = 2023,
     number = {10}
}

@article{Krauss:1983zn,
    author = "Krauss, Lawrence M. and Glashow, Sheldon L. and Schramm, David N.",
    title = "{Anti-neutrinos Astronomy and Geophysics}",
    reportNumber = "HUTP-83/A076",
    doi = "10.1038/310191a0",
    journal = "Nature",
    volume = "310",
    pages = "191--198",
    year = "1984"
}

@article{Salpeter:1955it,
    author = "Salpeter, Edwin E.",
    title = "{The Luminosity function and stellar evolution}",
    doi = "10.1086/145971",
    journal = "Astrophys. J.",
    volume = "121",
    pages = "161--167",
    year = "1955"
}

@article{Hopkins:2006bw,
    author = "Hopkins, Andrew M. and Beacom, John F.",
    title = "{On the normalisation of the cosmic star formation history}",
    eprint = "astro-ph/0601463",
    archivePrefix = "arXiv",
    doi = "10.1086/506610",
    journal = "Astrophys. J.",
    volume = "651",
    pages = "142--154",
    year = "2006"
}

@article{Horiuchi:2008jz,
    author = "Horiuchi, Shunsaku and Beacom, John F. and Dwek, Eli",
    title = "{The Diffuse Supernova Neutrino Background is detectable in Super-Kamiokande}",
    eprint = "0812.3157",
    archivePrefix = "arXiv",
    primaryClass = "astro-ph",
    doi = "10.1103/PhysRevD.79.083013",
    journal = "Phys. Rev. D",
    volume = "79",
    pages = "083013",
    year = "2009"
}

@article{deSalas:2020pgw,
    author = "de Salas, P. F. and Forero, D. V. and Gariazzo, S. and Mart\'\i{}nez-Mirav\'e, P. and Mena, O. and Ternes, C. A. and T\'ortola, M. and Valle, J. W. F.",
    title = "{2020 global reassessment of the neutrino oscillation picture}",
    eprint = "2006.11237",
    archivePrefix = "arXiv",
    primaryClass = "hep-ph",
    doi = "10.1007/JHEP02(2021)071",
    journal = "JHEP",
    volume = "02",
    pages = "071",
    year = "2021"
}

@article{Neustadt:2021a,
       author = {{Neustadt}, J.~M.~M. and {Kochanek}, C.~S. and {Stanek}, K.~Z. and {Basinger}, C. and {Jayasinghe}, T. and {Garling}, C.~T. and {Adams}, S.~M. and {Gerke}, J.},
        title = "{The search for failed supernovae with the Large Binocular Telescope: a new candidate and the failed SN fraction with 11 yr of data}",
      journal = {Monthly Notices of the Royal Astronomical Society},
         year = 2021,
        month = nov,
       volume = {508},
       number = {1},
        pages = {516-528},
          doi = {10.1093/mnras/stab2605},
archivePrefix = {arXiv},
       eprint = {2104.03318},
 primaryClass = {astro-ph.SR},
       adsurl = {https://ui.adsabs.harvard.edu/abs/2021MNRAS.508..516N}
}

@article{Gerke:2015a,
       author = {{Gerke}, J.~R. and {Kochanek}, C.~S. and {Stanek}, K.~Z.},
        title = "{The search for failed supernovae with the Large Binocular Telescope: first candidates}",
      journal = {Monthly Notices of the Royal Astronomical Society},
         year = 2015,
        month = jul,
       volume = {450},
       number = {3},
        pages = {3289-3305},
          doi = {10.1093/mnras/stv776},
archivePrefix = {arXiv},
       eprint = {1411.1761},
 primaryClass = {astro-ph.SR},
       adsurl = {https://ui.adsabs.harvard.edu/abs/2015MNRAS.450.3289G}
}

@article{Adams:2017a,
       author = {{Adams}, S.~M. and {Kochanek}, C.~S. and {Gerke}, J.~R. and {Stanek}, K.~Z.},
        title = "{The search for failed supernovae with the Large Binocular Telescope: constraints from 7 yr of data}",
      journal = {Monthly Notices of the Royal Astronomical Society},
         year = 2017,
        month = aug,
       volume = {469},
       number = {2},
        pages = {1445-1455},
          doi = {10.1093/mnras/stx898},
archivePrefix = {arXiv},
       eprint = {1610.02402},
 primaryClass = {astro-ph.SR},
       adsurl = {https://ui.adsabs.harvard.edu/abs/2017MNRAS.469.1445A}
}

@article{Kresse:2020nto,
    author = "Kresse, Daniel and Ertl, Thomas and Janka, Hans-Thomas",
    title = "{Stellar Collapse Diversity and the Diffuse Supernova Neutrino Background}",
    eprint = "2010.04728",
    archivePrefix = "arXiv",
    primaryClass = "astro-ph.HE",
    doi = "10.3847/1538-4357/abd54e",
    journal = "Astrophys. J.",
    volume = "909",
    number = "2",
    pages = "169",
    year = "2021"
}

@article{Planck:2018vyg,
    author = "Aghanim, N. and others",
    collaboration = "Planck",
    title = "{Planck 2018 results. VI. Cosmological parameters}",
    eprint = "1807.06209",
    archivePrefix = "arXiv",
    primaryClass = "astro-ph.CO",
    doi = "10.1051/0004-6361/201833910",
    journal = "Astron. Astrophys.",
    volume = "641",
    pages = "A6",
    year = "2020",
    note = "[Erratum: Astron.Astrophys. 652, C4 (2021)]"
}

@misc{Garching,
    author = "",
    title = "Garching Core-Collapse Supernova Archive",
    url = "https://wwwmpa.mpa-garching.mpg.de/ccsnarchive/" 
}

@article{Lattimer:1991nc,
    author = "Lattimer, James M. and Swesty, F. Douglas",
    title = "{A Generalized equation of state for hot, dense matter}",
    doi = "10.1016/0375-9474(91)90452-C",
    journal = "Nucl. Phys. A",
    volume = "535",
    pages = "331--376",
    year = "1991"
}

@article{Ando:2004hc,
    author = "Ando, Shin'ichiro and Sato, Katsuhiko",
    title = "{Relic neutrino background from cosmological supernovae}",
    eprint = "astro-ph/0410061",
    archivePrefix = "arXiv",
    doi = "10.1088/1367-2630/6/1/170",
    journal = "New J. Phys.",
    volume = "6",
    pages = "170",
    year = "2004"
}

@article{SNO:2004eru,
    author = "Aharmim, B. and others",
    collaboration = "SNO",
    title = "{Electron antineutrino search at the Sudbury Neutrino Observatory}",
    eprint = "hep-ex/0407029",
    archivePrefix = "arXiv",
    doi = "10.1103/PhysRevD.70.093014",
    journal = "Phys. Rev. D",
    volume = "70",
    pages = "093014",
    year = "2004"
}

@article{Borexino:2019wln,
    author = "Agostini, M. and others",
    collaboration = "Borexino",
    title = "{Search for low-energy neutrinos from astrophysical sources with Borexino}",
    eprint = "1909.02422",
    archivePrefix = "arXiv",
    primaryClass = "hep-ex",
    reportNumber = "FERMILAB-PUB-21-152-AE",
    doi = "10.1016/j.astropartphys.2020.102509",
    journal = "Astropart. Phys.",
    volume = "125",
    pages = "102509",
    year = "2021"
}

@article{KamLAND:2021gvi,
    author = "Abe, S. and others",
    collaboration = "KamLAND",
    title = "{Limits on Astrophysical Antineutrinos with the KamLAND Experiment}",
    eprint = "2108.08527",
    archivePrefix = "arXiv",
    primaryClass = "astro-ph.HE",
    doi = "10.3847/1538-4357/ac32c1",
    journal = "Astrophys. J.",
    volume = "925",
    number = "1",
    pages = "14",
    year = "2022"
}

@article{Super-Kamiokande:2020frs,
    author = "Abe, K. and others",
    collaboration = "Super-Kamiokande",
    title = "{Search for solar electron anti-neutrinos due to spin-flavor precession in the Sun with Super-Kamiokande-IV}",
    eprint = "2012.03807",
    archivePrefix = "arXiv",
    primaryClass = "hep-ex",
    doi = "10.1016/j.astropartphys.2022.102702",
    journal = "Astropart. Phys.",
    volume = "139",
    pages = "102702",
    year = "2022"
}

@unpublished{SK-Gd:2024,
  location = {{Milano, Italy}},
  title = {},
  url = {https://agenda.infn.it/event/37867/contributions/233922/attachments/122065/178295/20240620_v9.pdf},
  langid = {english},
  type = {Talk},
  howpublished = {Talk},
  eventtitle = {XXXI International Conference on Neutrino Physics and Astrophysics},
  author = {Masayuki Harada},
  date = {June 2024},
  note = "{Talk ``Review of Diffuse supernova Neutrino Background'' at the XXXI International Conference on Neutrino Physics and Astrophysics (Neutrino 2024), Milano (Italy), June 16-22, 2024.}",
}

@article{Aguilera-Dena:2018ork,
    author = "Aguilera-Dena, David R. and Langer, Norbert and Moriya, Takashi J. and Schootemeijer, Abel",
    title = "{Related progenitor models for long-duration gamma ray bursts and Type Ic superluminous supernovae}",
    eprint = "1804.07317",
    archivePrefix = "arXiv",
    primaryClass = "astro-ph.SR",
    doi = "10.3847/1538-4357/aabfc1",
    journal = "Astrophys. J.",
    volume = "858",
    number = "2",
    pages = "115",
    year = "2018"
}

@article{Frohmaier:2020pec,
    author = "Frohmaier, C. and others",
    title = "{From core collapse to superluminous: The rates of massive stellar explosions from the Palomar Transient Factory}",
    eprint = "2010.15270",
    archivePrefix = "arXiv",
    primaryClass = "astro-ph.HE",
    doi = "10.1093/mnras/staa3607",
    journal = "Mon. Not. Roy. Astron. Soc.",
    volume = "500",
    number = "4",
    pages = "5142--5158",
    year = "2020"
}

@article{JUNO:2022lpc,
    author = "Abusleme, Angel and others",
    collaboration = "JUNO",
    title = "{Prospects for detecting the diffuse supernova neutrino background with JUNO}",
    eprint = "2205.08830",
    archivePrefix = "arXiv",
    primaryClass = "hep-ex",
    doi = "10.1088/1475-7516/2022/10/033",
    journal = "JCAP",
    volume = "10",
    pages = "033",
    year = "2022"
}

@article{Martinez-Mirave:2024hfd,
    author = "Mart\'\i{}nez-Mirav\'e, Pablo and Tamborra, Irene and T\'ortola, Mariam",
    title = "{The Sun and core-collapse supernovae are leading probes of the neutrino lifetime}",
    eprint = "2402.00116",
    archivePrefix = "arXiv",
    primaryClass = "astro-ph.HE",
    doi = "10.1088/1475-7516/2024/05/002",
    journal = "JCAP",
    volume = "05",
    pages = "002",
    year = "2024"
}

@article{Moller:2018kpn,
    author = "M\o{}ller, Klaes and Suliga, Anna M. and Tamborra, Irene and Denton, Peter B.",
    title = "{Measuring the supernova unknowns at the next-generation neutrino telescopes through the diffuse neutrino background}",
    eprint = "1804.03157",
    archivePrefix = "arXiv",
    primaryClass = "astro-ph.HE",
    doi = "10.1088/1475-7516/2018/05/066",
    journal = "JCAP",
    volume = "05",
    pages = "066",
    year = "2018"
}

@article{Nakazato:2015rya,
    author = "Nakazato, Ken'ichiro and Mochida, Eri and Niino, Yuu and Suzuki, Hideyuki",
    title = "{Spectrum of the Supernova Relic Neutrino Background and Metallicity Evolution of Galaxies}",
    eprint = "1503.01236",
    archivePrefix = "arXiv",
    primaryClass = "astro-ph.HE",
    doi = "10.1088/0004-637X/804/1/75",
    journal = "Astrophys. J.",
    volume = "804",
    number = "1",
    pages = "75",
    year = "2015"
}

@article{Ashida:2023heb,
    author = "Ashida, Yosuke and Nakazato, Ken'ichiro and Tsujimoto, Takuji",
    title = "{Diffuse Neutrino Flux Based on the Rates of Core-collapse Supernovae and Black Hole Formation Deduced from a Novel Galactic Chemical Evolution Model}",
    eprint = "2305.13543",
    archivePrefix = "arXiv",
    primaryClass = "astro-ph.HE",
    doi = "10.3847/1538-4357/ace3ba",
    journal = "Astrophys. J.",
    volume = "953",
    number = "2",
    pages = "151",
    year = "2023"
}

@article{Ekanger:2023qzw,
    author = "Ekanger, Nick and Horiuchi, Shunsaku and Nagakura, Hiroki and Reitz, Samantha",
    title = "{Diffuse supernova neutrino background with up-to-date star formation rate measurements and long-term multidimensional supernova simulations}",
    eprint = "2310.15254",
    archivePrefix = "arXiv",
    primaryClass = "astro-ph.HE",
    doi = "10.1103/PhysRevD.109.023024",
    journal = "Phys. Rev. D",
    volume = "109",
    number = "2",
    pages = "023024",
    year = "2024"
}

@article{Anandagoda:2023sbg,
    author = "Anandagoda, Samalka and Hartmann, Dieter H. and Fryer, Christopher L. and Ajello, Marco and Desai, Abhishek and Hungerford, Aimee L. and The, Lih-Sin",
    title = "{Cosmic Supernova Neutrino and Gamma-Ray Backgrounds in the MeV Regime}",
    doi = "10.3847/1538-4357/acc84f",
    journal = "Astrophys. J.",
    volume = "950",
    number = "1",
    pages = "29",
    year = "2023"
}

@article{Ziegler:2022ivq,
    author = "Ziegler, Joshua J. and Edwards, Thomas D. P. and Suliga, Anna M. and Tamborra, Irene and Horiuchi, Shunsaku and Ando, Shin'ichiro and Freese, Katherine",
    title = "{Non-universal stellar initial mass functions: large uncertainties in star formation rates at z \ensuremath{\approx} 2\textendash{}4 and other astrophysical probes}",
    eprint = "2205.07845",
    archivePrefix = "arXiv",
    primaryClass = "astro-ph.GA",
    reportNumber = "UTTG-05-2022, N3AS-22-007, NORDITA 2022-026",
    doi = "10.1093/mnras/stac2748",
    journal = "Mon. Not. Roy. Astron. Soc.",
    volume = "517",
    number = "2",
    pages = "2471--2484",
    year = "2022"
}

@article{Horiuchi:2020jnc,
    author = "Horiuchi, Shunsaku and Kinugawa, Tomoya and Takiwaki, Tomoya and Takahashi, Koh and Kotake, Kei",
    title = "{Impact of binary interactions on the diffuse supernova neutrino background}",
    eprint = "2012.08524",
    archivePrefix = "arXiv",
    primaryClass = "astro-ph.HE",
    doi = "10.1103/PhysRevD.103.043003",
    journal = "Phys. Rev. D",
    volume = "103",
    number = "4",
    pages = "043003",
    year = "2021"
}

@article{Nagele:2021kvw,
    author = "Nagele, Chris and Umeda, Hideyuki and Takahashi, Koh and Yoshida, Takashi and Sumiyoshi, Kohsuke",
    title = "{Neutrino emission from the collapse of \ensuremath{\sim}104 M\ensuremath{\odot} Population III supermassive stars}",
    eprint = "2107.01761",
    archivePrefix = "arXiv",
    primaryClass = "astro-ph.HE",
    doi = "10.1093/mnras/stab2592",
    journal = "Mon. Not. Roy. Astron. Soc.",
    volume = "508",
    number = "1",
    pages = "828--841",
    year = "2021"
}

@article{Suwa:2008sf,
    author = "Suwa, Yudai and Takiwaki, Tomoya and Kotake, Kei and Sato, Katsuhiko",
    title = "{Impact of Rotation on Neutrino Emission and Relic Neutrino Background from Population III Stars}",
    eprint = "0806.1072",
    archivePrefix = "arXiv",
    primaryClass = "astro-ph",
    reportNumber = "UTAP-597, RESCEU-10-08",
    doi = "10.1088/0004-637X/690/1/913",
    journal = "Astrophys. J.",
    volume = "690",
    pages = "913--922",
    year = "2009"
}

@article{Nakazato:2005ek,
    author = "Nakazato, Ken'ichiro and Sumiyoshi, Kohsuke and Yamada, Shoichi",
    title = "{Gravitational collapse and neutrino emission of population III massive stars}",
    eprint = "astro-ph/0509868",
    archivePrefix = "arXiv",
    doi = "10.1086/504282",
    journal = "Astrophys. J.",
    volume = "645",
    pages = "519--533",
    year = "2006"
}

@article{Vitagliano:2019yzm,
    author = "Vitagliano, Edoardo and Tamborra, Irene and Raffelt, Georg G.",
    title = "{Grand Unified Neutrino Spectrum at Earth: Sources and Spectral Components}",
    eprint = "1910.11878",
    archivePrefix = "arXiv",
    primaryClass = "astro-ph.HE",
    reportNumber = "MPP-2019-205",
    doi = "10.1103/RevModPhys.92.045006",
    journal = "Rev. Mod. Phys.",
    volume = "92",
    pages = "45006",
    year = "2020"
}

@article{Pattavina:2020cqc,
    author = "Pattavina, Luca and Ferreiro Iachellini, Nahuel and Tamborra, Irene",
    title = "{Neutrino observatory based on archaeological lead}",
    eprint = "2004.06936",
    archivePrefix = "arXiv",
    primaryClass = "astro-ph.HE",
    doi = "10.1103/PhysRevD.102.063001",
    journal = "Phys. Rev. D",
    volume = "102",
    number = "6",
    pages = "063001",
    year = "2020"
}

@article{Lang:2016zhv,
    author = "Lang, Rafael F. and McCabe, Christopher and Reichard, Shayne and Selvi, Marco and Tamborra, Irene",
    title = "{Supernova neutrino physics with xenon dark matter detectors: A timely perspective}",
    eprint = "1606.09243",
    archivePrefix = "arXiv",
    primaryClass = "astro-ph.HE",
    doi = "10.1103/PhysRevD.94.103009",
    journal = "Phys. Rev. D",
    volume = "94",
    number = "10",
    pages = "103009",
    year = "2016"
}

@article{DUNE:2020lwj,
    author = "Abi, Babak and others",
    collaboration = "DUNE",
    title = "{Deep Underground Neutrino Experiment (DUNE), Far Detector Technical Design Report, Volume I Introduction to DUNE}",
    eprint = "2002.02967",
    archivePrefix = "arXiv",
    primaryClass = "physics.ins-det",
    reportNumber = "FERMILAB-PUB-20-024-ND, FERMILAB-DESIGN-2020-01",
    doi = "10.1088/1748-0221/15/08/T08008",
    journal = "JINST",
    volume = "15",
    number = "08",
    pages = "T08008",
    year = "2020"
}

@article{Schilbach:2018bsg,
    author = "Schilbach, T. S. H. and Caballero, O. L. and McLaughlin, G. C.",
    title = "{Black Hole Accretion Disk Diffuse Neutrino Background}",
    eprint = "1808.03627",
    archivePrefix = "arXiv",
    primaryClass = "astro-ph.HE",
    doi = "10.1103/PhysRevD.100.043008",
    journal = "Phys. Rev. D",
    volume = "100",
    number = "4",
    pages = "043008",
    year = "2019"
}

@article{Wei:2024qgy,
    author = "Wei, Yun-Feng and Liu, Tong and Song, Cui-Ying",
    title = "{Contribution of Neutrino-dominated Accretion Flows to the Cosmic MeV Neutrino Background}",
    eprint = "2403.16856",
    archivePrefix = "arXiv",
    primaryClass = "astro-ph.HE",
    doi = "10.3847/1538-4357/ad3824",
    journal = "Astrophys. J.",
    volume = "966",
    number = "1",
    pages = "101",
    year = "2024"
}

@article{Lunardini:2012ne,
    author = "Lunardini, Cecilia and Tamborra, Irene",
    title = "{Diffuse supernova neutrinos: oscillation effects, stellar cooling and progenitor mass dependence}",
    eprint = "1205.6292",
    archivePrefix = "arXiv",
    primaryClass = "astro-ph.SR",
    reportNumber = "MPP-2012-90",
    doi = "10.1088/1475-7516/2012/07/012",
    journal = "JCAP",
    volume = "07",
    pages = "012",
    year = "2012"
}

@article{Lunardini:2009ya,
    author = "Lunardini, Cecilia",
    title = "{Diffuse neutrino flux from failed supernovae}",
    eprint = "0901.0568",
    archivePrefix = "arXiv",
    primaryClass = "astro-ph.SR",
    doi = "10.1103/PhysRevLett.102.231101",
    journal = "Phys. Rev. Lett.",
    volume = "102",
    pages = "231101",
    year = "2009"
}

@article{Nakazato:2024gem,
    author = "Nakazato, Ken'ichiro and Akaho, Ryuichiro and Ashida, Yosuke and Tsujimoto, Takuji",
    title = "{Impacts of Black-Hole-Forming Supernova Explosions on the Diffuse Neutrino Background}",
    journal="",
    eprint = "2406.13276",
    archivePrefix = "arXiv",
    primaryClass = "astro-ph.HE",
    month = "6",
    year = "2024"
}

@article{Super-Kamiokande:2021the,
    author = "Abe, K. and others",
    collaboration = "Super-Kamiokande",
    title = "{First gadolinium loading to Super-Kamiokande}",
    eprint = "2109.00360",
    archivePrefix = "arXiv",
    primaryClass = "physics.ins-det",
    doi = "10.1016/j.nima.2021.166248",
    journal = "Nucl. Instrum. Meth. A",
    volume = "1027",
    pages = "166248",
    year = "2022"
}

@article{Super-Kamiokande:2024kcb,
    author = "Abe, K. and others",
    collaboration = "Super-Kamiokande",
    title = "{Second gadolinium loading to Super-Kamiokande}",
    eprint = "2403.07796",
    archivePrefix = "arXiv",
    primaryClass = "physics.ins-det",
    doi = "10.1016/j.nima.2024.169480",
    journal = "Nucl. Instrum. Meth. A",
    volume = "1065",
    pages = "169480",
    year = "2024"
}

@article{Suliga:2021hek,
    author = "Suliga, Anna M. and Beacom, John F. and Tamborra, Irene",
    title = "{Towards probing the diffuse supernova neutrino background in all flavors}",
    eprint = "2112.09168",
    archivePrefix = "arXiv",
    primaryClass = "astro-ph.HE",
    reportNumber = "N3AS-21-016",
    doi = "10.1103/PhysRevD.105.043008",
    journal = "Phys. Rev. D",
    volume = "105",
    number = "4",
    pages = "043008",
    year = "2022"
}

@article{Strumia:2003zx,
    author = "Strumia, Alessandro and Vissani, Francesco",
    title = "{Precise quasielastic neutrino/nucleon cross-section}",
    eprint = "astro-ph/0302055",
    archivePrefix = "arXiv",
    reportNumber = "IFUP-TH-2003-2",
    doi = "10.1016/S0370-2693(03)00616-6",
    journal = "Phys. Lett. B",
    volume = "564",
    pages = "42--54",
    year = "2003"
}

@article{Akhmedov:2022txm,
    author = "Akhmedov, Evgeny and Mart\'\i{}nez-Mirav\'e, Pablo",
    title = "{Solar $ {\overline{\nu}}_e $ flux: revisiting bounds on neutrino magnetic moments and solar magnetic field}",
    eprint = "2207.04516",
    archivePrefix = "arXiv",
    primaryClass = "hep-ph",
    doi = "10.1007/JHEP10(2022)144",
    journal = "JHEP",
    volume = "10",
    pages = "144",
    year = "2022"
}

@article{Maksimovic:2021dmz,
    author = "Maksimovi\'c, David and Nieslony, Michael and Wurm, Michael",
    title = "{CNNs for enhanced background discrimination in DSNB searches in large-scale water-Gd detectors}",
    eprint = "2104.13426",
    archivePrefix = "arXiv",
    primaryClass = "physics.ins-det",
    doi = "10.1088/1475-7516/2021/11/051",
    journal = "JCAP",
    volume = "11",
    number = "11",
    pages = "051",
    year = "2021"
}

@phdthesis{Kunxian:2016joi,
    author = "Kunxian, Huang",
    title = "{Measurement of the Neutrino-Oxygen Neutral Current Quasi-elastic Interaction Cross-section by Observing Nuclear De-excitation $\gamma$-rays in the T2K Experiment}",
    doi = "10.14989/doctor.k19501",
    school = "Kyoto U.",
    year = "2016"
}

@phdthesis{Ashida:2020erk,
    author = "Ashida, Yosuke",
    title = "{Measurement of Neutrino and Antineutrino Neutral-Current Quasielastic-like Interactions and Applications to Supernova Relic Neutrino Searches}",
    school = "Kyoto U.",
    year = "2020"
}

@article{Hyper-Kamiokande:2018ofw,
    author = "Abe, K. and others",
    collaboration = "Hyper-Kamiokande",
    title = "{Hyper-Kamiokande Design Report}",
    journal ="",
    eprint = "1805.04163",
    archivePrefix = "arXiv",
    primaryClass = "physics.ins-det",
    month = "5",
    year = "2018"
}

@article{Hyper-Kamiokande:2016srs,
    author = "Abe, K. and others",
    collaboration = "Hyper-Kamiokande",
    title = "{Physics potentials with the second Hyper-Kamiokande detector in Korea}",
    eprint = "1611.06118",
    archivePrefix = "arXiv",
    primaryClass = "hep-ex",
    doi = "10.1093/ptep/pty044",
    journal = "PTEP",
    volume = "2018",
    number = "6",
    pages = "063C01",
    year = "2018"
}

@article{DUNE:2020ypp,
    author = "Abi, Babak and others",
    collaboration = "DUNE",
    title = "{Deep Underground Neutrino Experiment (DUNE), Far Detector Technical Design Report, Volume II: DUNE Physics}",
    journal="",
    eprint = "2002.03005",
    archivePrefix = "arXiv",
    primaryClass = "hep-ex",
    reportNumber = "FERMILAB-PUB-20-025-ND, FERMILAB-DESIGN-2020-02",
    month = "2",
    year = "2020"
}

@article{JUNO:2021vlw,
    author = "Abusleme, Angel and others",
    collaboration = "JUNO",
    title = "{JUNO physics and detector}",
    eprint = "2104.02565",
    archivePrefix = "arXiv",
    primaryClass = "hep-ex",
    doi = "10.1016/j.ppnp.2021.103927",
    journal = "Prog. Part. Nucl. Phys.",
    volume = "123",
    pages = "103927",
    year = "2022"
}

@article{Simpson:2018snj,
    author = "Simpson, Charles",
    collaboration = "Super-Kamiokande",
    title = "{Physics Potential of Super-K Gd}",
    doi = "10.22323/1.340.0008",
    journal = "PoS",
    volume = "ICHEP2018",
    pages = "008",
    year = "2019"
}

@article{Ricciardi:2022pru,
    author = "Ricciardi, Giulia and Vignaroli, Natascia and Vissani, Francesco",
    title = "{An accurate evaluation of electron (anti-)neutrino scattering on nucleons}",
    eprint = "2206.05567",
    archivePrefix = "arXiv",
    primaryClass = "hep-ph",
    doi = "10.1007/JHEP08(2022)212",
    journal = "JHEP",
    volume = "08",
    pages = "212",
    year = "2022"
}

@inproceedings{Stock:2024tmd,
    author = "Stock, Matthias Raphael",
    collaboration = "JUNO",
    title = "{Status and Prospects of the JUNO Experiment}",
    booktitle = "{17th International Workshop on Tau Lepton Physics}",
    eprint = "2405.07321",
    archivePrefix = "arXiv",
    primaryClass = "physics.ins-det",
    month = "5",
    year = "2024"
}

@article{Abdalla:2022yfr,
    author = "Abdalla, Elcio and others",
    title = "{Cosmology intertwined: A review of the particle physics, astrophysics, and cosmology associated with the cosmological tensions and anomalies}",
    eprint = "2203.06142",
    archivePrefix = "arXiv",
    primaryClass = "astro-ph.CO",
    reportNumber = "FERMILAB-CONF-22-192-SCD",
    doi = "10.1016/j.jheap.2022.04.002",
    journal = "JHEAp",
    volume = "34",
    pages = "49--211",
    year = "2022"
}

@article{Marti:2019dof,
    author = "Marti, Ll. and others",
    title = "{Evaluation of gadolinium\textquoteright{}s action on water Cherenkov detector systems with EGADS}",
    eprint = "1908.11532",
    archivePrefix = "arXiv",
    primaryClass = "physics.ins-det",
    doi = "10.1016/j.nima.2020.163549",
    journal = "Nucl. Instrum. Meth. A",
    volume = "959",
    pages = "163549",
    year = "2020"
}

@article{Li:2022myd,
    author = "Li, Yu-Feng and Vagins, Mark and Wurm, Michael",
    title = "{Prospects for the Detection of the Diffuse Supernova Neutrino Background with the Experiments SK-Gd and JUNO}",
    eprint = "2201.12920",
    archivePrefix = "arXiv",
    primaryClass = "astro-ph.HE",
    doi = "10.3390/universe8030181",
    journal = "Universe",
    volume = "8",
    number = "3",
    pages = "181",
    year = "2022"
}

@article{Chabrier_2014,
   title={VARIATIONS OF THE STELLAR INITIAL MASS FUNCTION IN THE PROGENITORS OF MASSIVE EARLY-TYPE GALAXIES AND IN EXTREME STARBURST ENVIRONMENTS},
   volume={796},
   ISSN={1538-4357},
   url={http://dx.doi.org/10.1088/0004-637X/796/2/75},
   DOI={10.1088/0004-637x/796/2/75},
   number={2},
   journal={The Astrophysical Journal},
   publisher={American Astronomical Society},
   author={Chabrier, Gilles and Hennebelle, Patrick and Charlot, Stéphane},
   year={2014},
   month=nov, pages={75} }

@article{Aloy:2020svw,
    author = "Aloy, Miguel-{\'A}ngel and Obergaulinger, Martin",
    title = "{Magnetorotational core collapse of possible GRB progenitors \textendash{} II. Formation of protomagnetars and collapsars}",
    eprint = "2008.03779",
    archivePrefix = "arXiv",
    primaryClass = "astro-ph.HE",
    doi = "10.1093/mnras/staa3273",
    journal = "Mon. Not. Roy. Astron. Soc.",
    volume = "500",
    number = "4",
    pages = "4365--4397",
    year = "2020"
}

@article{Obergaulinger:2020cqq,
    author = "Obergaulinger, Martin and Aloy, Miguel {\'A}ngel",
    title = "{Magnetorotational core collapse of possible GRB progenitors. III. Three-dimensional models}",
    eprint = "2008.07205",
    archivePrefix = "arXiv",
    primaryClass = "astro-ph.HE",
    doi = "10.1093/mnras/stab295",
    journal = "Mon. Not. Roy. Astron. Soc.",
    volume = "503",
    number = "4",
    pages = "4942--4963",
    year = "2021"
}

@article{Lipunova:2009qf,
    author = "Lipunova, G. V. and Gorbovskoy, E. S. and Bogomazov, A. I. and Lipunov, V. M.",
    title = "{Population synthesis of gamma-ray bursts with precursor activity and the spinar paradigm}",
    eprint = "0903.3169",
    archivePrefix = "arXiv",
    primaryClass = "astro-ph.HE",
    doi = "10.1111/j.1365-2966.2009.15079.x",
    journal = "Mon. Not. Roy. Astron. Soc.",
    volume = "397",
    pages = "1695--1704",
    year = "2009"
}

@article{Obergaulinger:2017qno,
    author = "Obergaulinger, Martin and Aloy, Miguel {\'A}ngel",
    title = "{Protomagnetar and black hole formation in high-mass stars}",
    eprint = "1703.09893",
    archivePrefix = "arXiv",
    primaryClass = "astro-ph.SR",
    doi = "10.1093/mnrasl/slx046",
    journal = "Mon. Not. Roy. Astron. Soc.",
    volume = "469",
    number = "1",
    pages = "L43--L47",
    year = "2017"
}

@article{DUNE:2024wvj,
    author = "Abed Abud, Adam and others",
    collaboration = "DUNE",
    journal ="",
    title = "{DUNE Phase II: Scientific Opportunities, Detector Concepts, Technological Solutions}",
    eprint = "2408.12725",
    archivePrefix = "arXiv",
    primaryClass = "physics.ins-det",
    reportNumber = "FERMILAB-TM-2833-LBNF",
    month = "8",
    year = "2024"
}

\end{document}